\documentclass[oldversion]{aa}  
\usepackage{graphicx,amsmath}
\usepackage{amssymb}
\usepackage{color}
\usepackage{natbib}             
\usepackage{url}
\usepackage{multirow}
\usepackage{threeparttable}
\bibpunct{(}{)}{;}{a}{}{,} 
\usepackage{txfonts}
\usepackage{subfig}
\usepackage[normalem]{ulem}
\usepackage{deluxetable}

\makeatletter
\def\@to{to}
\makeatother

\newcommand{\Mwater}{$m_{\rm water}$\,}

\newcommand{\rc}{$r_{\rm core}$\,}
\newcommand{\rsolid}{$r_{\rm core+mantle}$\,}

\newcommand{\fesi}{{\rm Fe}/{\rm Si}_{\rm bulk}\,}
\newcommand{\mgsi}{{\rm Mg}/{\rm Si}_{\rm bulk}\,}
\newcommand{\fesistar}{{\rm Fe}/{\rm Si}_{\rm star}\,}
\newcommand{\mgsistar}{{\rm Mg}/{\rm Si}_{\rm star}\,}
\newcommand{\fesima}{{\rm Fe}/{\rm Si}_{\rm mantle}\,}
\newcommand{\mgsima}{{\rm Mg}/{\rm Si}_{\rm mantle}\,}

\usepackage[normalem]{ulem}
\usepackage{color}
\definecolor{red}{RGB}{255,0,0}

\def\approxinf{%
  \def\p{%
    \setbox0=\vbox{\hbox{$<$}}%
    \ht0=0.6ex \box0 }%
  \def\s{%
    \vbox{\hbox{$\sim$}}%
  }%
  \mathrel{\raisebox{0.7ex}{%
      \mbox{$\underset{\s}{\p}$}%
    }}%
}

\def\ms{\hbox{\,m\,s$^{-1}$}}         
\def\m2s2{\hbox{\,m$^{2}$\,s$^{-2}$}} 
\def\kms{\hbox{\,km\,s$^{-1}$}}       

\begin{document}

   \title{The 55\,Cnc system reassessed\thanks{Individual APT photometric measurements for 55\,Cnc and its comparison stars, as well as RV measurements of 55\,Cnc, are available at the CDS via anonymous ftp to cdsarc.u-strasbg.fr (130.79.128.5) or via http://cdsarc.u-strasbg.fr/viz-bin/qcat?J/A+A/}}
   
   \author{
   V.~Bourrier\inst{1},
   X.~Dumusque\inst{1}, 
   C.~Dorn\inst{2},
   G.W.~Henry\inst{3},
   N.~Astudillo-Defru\inst{4},
   J.~Rey\inst{1}, 
   B.~Benneke\inst{5}, 
   G.~H\'ebrard\inst{6,7},
   C.~Lovis\inst{1},  
   B.O.~Demory\inst{8}, 
   C.~Moutou\inst{9,10},
   D.~Ehrenreich\inst{1}   
        }

\authorrunning{V.~Bourrier et al.}
\titlerunning{The 55\,Cnc system reassessed}

\offprints{V.B. (\email{vincent.bourrier@unige.ch})}

\institute{
Observatoire de l'Universit\'e de Gen\`eve, 51 chemin des Maillettes, 1290 Sauverny, Switzerland
\and 
University of Zurich, Institut of Computational Sciences, University of Zurich, Winterthurerstrasse 190, CH-8057, Zurich, Switzerland.
\and 
Center of Excellence in Information Systems, Tennessee State University, Nashville, Tennessee 37209, USA. 
\and
Universidad de Concepci\'on, Departamento de Astronom\'ia, Casilla 160-C, Concepci\'on, Chile 
\and 
D\'epartement de Physique, Universit\'e de Montr\'eal, Montreal, H3T J4, Canada
\and 
CNRS, UMR 7095, Institut d’Astrophysique de Paris, 98bis boulevard Arago, F-75014 Paris, France
\and 
UPMC Univ. Paris 6, UMR 7095, Institut d’Astrophysique de Paris, 98bis boulevard Arago, F-75014 Paris, France
\and 
  University of Bern, Center for Space and Habitability, Sidlerstrasse 5, CH-3012 Bern, Switzerland.
\and 
CFHT/CNRS, 65-1238 Mamalahoa Highway, Kamuela HI 96743, USA
\and
Aix Marseille Univ, CNRS, LAM, Laboratoire d'Astrophysique de Marseille, Marseille, France    
}

   \date{} 
 
  \abstract
{Orbiting a bright, nearby star the 55\,Cnc system offers a rare opportunity to study a multiplanet system that has a wide range of planetary masses and orbital distances. Using two decades of photometry and spectroscopy data, we have measured the rotation of the host star and its solar-like magnetic cycle. Accounting for this cycle in our velocimetric analysis of the system allows us to revise the properties of the outermost giant planet and its four planetary companions. The innermost planet 55\,Cnc\,e is an unusually close-in super-Earth, whose transits have allowed for detailed follow-up studies. Recent observations favor the presence of a substantial atmosphere yet its composition, and the nature of the planet, remain unknown. We combined our derived planet mass ($M_\mathrm{p}$ = 8.0$\pm$0.3\,M$_{\rm Earth}$) with refined measurement of its optical radius derived from HST/STIS observations ($R_\mathrm{p}$ = 1.88$\pm$0.03\,R$_{\rm Earth}$ over 530-750\,nm) to revise the density of 55\,Cnc\,e ($\rho$ = 6.7$\pm$0.4\,g\,cm$^{-3}$). Based on these revised properties we have characterized possible interiors of 55\,Cnc\,e using a generalized Bayesian model. We confirm that the planet is likely surrounded by a heavyweight atmosphere, contributing a few percents of the planet radius. While we cannot exclude the presence of a water layer underneath the atmosphere, this scenario is unlikely given the observations of the planet across the entire spectrum and its strong irradiation. Follow-up observations of the system in photometry and in spectroscopy over different time-scales are needed to further investigate the nature and origin of this iconic super-Earth.}

\keywords{planetary systems - Stars: individual: 55\,Cnc}

   \maketitle

\section{Introduction}
\label{intro} 

Visible to the naked eye, the G8 dwarf 55\,Cnc (V=5.95, d=12.3\,pc, \citealt{vonbraun2011}) hosts a diverse system of at least five exoplanets (\citealt{Butler1997}, \citealt{Marcy2002}, \citealt{McArthur2004}, \citealt{Fischer2008}, \citealt{Dawson2010}), including a super-Earth orbiting in less than a day (55\,Cnc e), a warm Jupiter possibly at the limit of atmospheric stability (55\,Cnc b; \citealt{Ehrenreich2012}), and a gas giant with one of the longest known orbital periods (55\,Cnc d, $\sim$15\,years). 55\,Cnc, which is also in a binary system with an M dwarf at a projected separation of about 1060\,au (\citealt{Mugrauer2006}), is one of the three brightest stars known to host a transiting super-Earth (between HD\,219134, V=5.5, d=6.5\,pc, \citealt{Motalebi2015}; and HD\,97658, V=7.7, d=21.1\,pc, \citealt{Dragomir2013_HD976}). Radial velocity measurements and transit observations of 55\,Cnc\,e have refined its mass (8\,M$_\mathrm{Earth}$) and radius (1.9\,$R_\mathrm{Earth}$) over the years \citep[][]{Fischer2008,Dawson2010, Winn2011, demory2011, Demory2012, Gillon2012, Endl2012, Dragomir2013, Demory2016a, Fischer2017}, up to the point where its bulk density can be measured precisely enough to constrain its interior structure. 55\,Cnc\,e is one of the most massive members of the population of ultra-short period planets ($P\approxinf$1\,day) and stands on the upper radius side of the ``evaporation valley'' (\citealt{Fulton2017}) that might separate large super-Earths massive enough to retain H/He envelopes with mass fractions of a few percent, and small rocky super-Earths with atmospheres that contribute negligibly to their size. The study of 55\,Cnc\,e bulk and atmospheric composition is thus particularly important to our understanding of the formation and evolution of small, close-in planets. \\

The super-Earth 55\,Cnc\,e has been the focus of detailed studies from the ultraviolet to the mid-infrared, yet its nature remains shrouded in mystery. Transit observations have shown that the planet does not harbour a hydrogen exosphere (\citealt{Ehrenreich2012}) or an extended water-rich atmosphere (\citealt{Esteves2017}), while the peculiar shape of its infrared phase curve is consistent with a heavyweight atmosphere (e.g., dust, metals, or water) rather than a magmatic surface with no atmosphere  (\citealt{Demory2016b}, \citealt{Angelo2017}). High-resolution spectroscopy revealed changes in the transit depth of the optical sodium and singly-ionized calcium lines, possibly arising from variability in the structure of a putative exosphere (\citealt{RiddenHarper2016}), while stellar emission lines in the far-ultraviolet showed variations that could trace strong interactions between 55\,Cnc\,e and the stellar corona (Bourrier et al. 2018, submitted). Furthermore, Spitzer observations spanning three years have revealed significant temporal variability in the dayside-averaged thermal emission (\citealt{Demory2016a}), while {\it MOST} observations of the planet showed a significant decrease in the visible phase-curve amplitude between 2011 (\citealt{Winn2011}) and 2012 (\citealt{Dragomir2013}). All these elements point to the presence of a variable source of opacity in the atmosphere or at the surface of 55\,Cnc\,e. Multiwavelength observations over different timescales are required to determine the nature of this source and characterize its variability. Understanding its origin further requires that we constrain the interior structure and composition of 55\,Cnc\,e, which is the objective of the present study. \\
We investigated the activity of the star in Sect.~\ref{sec:stellar_prop}, and provide an updated velocimetric analysis of the planetary system in Sect.~\ref{sec:vel_ana} that includes the stellar magnetic cycle effect for the first time. Ground-based and space-borne transit observations of 55\,Cnc\,e are presented in Sect.~\ref{sec:photom_ana}, and we combine our revised mass and radius measurements of the planet to model its interior in Sect.~\ref{sec:intern_str}. We discuss the properties of 55\,Cnc e in Sect.~\ref{sec:disc}, and we draw concludions from this updated analysis on the 55\,Cnc planetary system in Sect.~\ref{sec:conc}.\\

\begin{table*}[tbh]
\caption{Properties of 55\,Cnc used and derived in our analysis.}\centering
\begin{threeparttable}
\begin{tabular}{lccll}
\hline
\noalign{\smallskip}  
\textbf{Parameter}      & \textbf{Symbol}                       &           \textbf{Value}                               & \textbf{Unit}         & \textbf{Reference} \\   
\noalign{\smallskip}
\hline
\hline
\noalign{\smallskip} 
Radius  & $R_{\star}$           &0.943$\pm$0.010                                                                                                                 &$R_{sun}$  & \citealt{vonbraun2011} \\
Mass            & $M_{\star}$                                                   & 0.905$\pm$0.015                  &$M_{sun}$   & \citealt{vonbraun2011}  \\
Effective temperature    & $T_{\rm eff}$                                        & 5172$\pm$18                                                                                                              &K & \citealt{Yee2017} \\ 
Surface gravity                   & $log_{10}(g)$ & 4.43$\pm$0.02 &  & \citealt{Yee2017} \\
Metallicity               & $[Fe/H]$ & 0.35$\pm$0.1 &  & \citealt{Yee2017} \\
Limb-darkening coefficients & $u_\mathrm{1}$ &  0.544$\pm$0.008                                                                                                  &   & This work\\
                                                        & $u_\mathrm{2}$ &       0.186$\pm$0.004                                                                                                  &   &   This work  \\
Cycle period & $P_\mathrm{mag}$ & 10.5$\pm$0.3  & years                                                                                                            & This work, derived from KECK HIRES S-index\\                                                                
Rotation period & $P_\mathrm{\star}$ &  38.8$\pm$0.05  &  days                                                                                                     & This work\\                                                 
Projected rotational velocity & $V$\,sin$i_\mathrm{\star}$ &    $<$1.23$\pm$0.01  &  km\,s$^{-1}$                                                                                                           & This work\\ 

\hline
\end{tabular}
\begin{tablenotes}[para,flushleft]
  \end{tablenotes}
  \end{threeparttable}
\label{table:tab_paramsfit}
\end{table*}


\section{Analysis of stellar activity}
\label{sec:stellar_prop}

\subsection{Magnetic cycle of 55\,Cnc}

We acquired 2243 good photometric observations of 55\,Cnc during 17 consecutive observing seasons between 2000 November 12 and 2017 April 17, all with the T8 0.80~m automatic photoelectric telescope (APT) at Fairborn Observatory in southern Arizona. The T8 APT is one of several automated telescopes operated at Fairborn by Tennessee State University and is equipped with a two-channel precision photometer that uses a dichroic filter and two EMI 9124QB bi-alkali photomultiplier tubes to separate and simultaneously count photons in the Str\"omgren $b$ and $y$ passbands \citep{Henry_1999}. 

We programmed the APT to make sequential brightness measurements of our program star 55~Cnc ($P$: $V=5.96$, $B-V=0.87$, G8V) along with the three comparison stars HD~76572 ($C1$: $V=6.25$, $B-V=0.47$, F6IV~V), HD~77190 ($C2$: $V=6.07$, 
$B-V=0.24$, A8V), and HD~79929 ($C3$: $V=6.77$, $B-V=0.41$, F6V).  From the raw counts in the two passbands, we computed six permutations of differential magnitudes from the four stars, namely, $P-C1_{by}$, $P-C2_{by}$, $P-C3_{by}$,
$C3-C2_{by}$, $C3-C1_{by}$, and $C2-C1_{by}$. We corrected the differential magnitudes for atmospheric extinction and transformed them to the Str\"omgren system. To improve our photometric precision, we combined the differential 
$b$ and $y$ observations into a single $(b+y)/2$ passband, as indicated above with the subscript $by$. Further information concerning our automated telescopes, precision photometers, and observing and data reduction techniques can be found in \citet{Henry_1995b,Henry_1995a}, \citet{Henry_1999} and \citet{Eaton2003} and references therein.

\begin{center}
\begin{figure}[h!]
\centering
\includegraphics[trim=0cm 7cm 0cm 7cm,clip=true,width=\columnwidth]{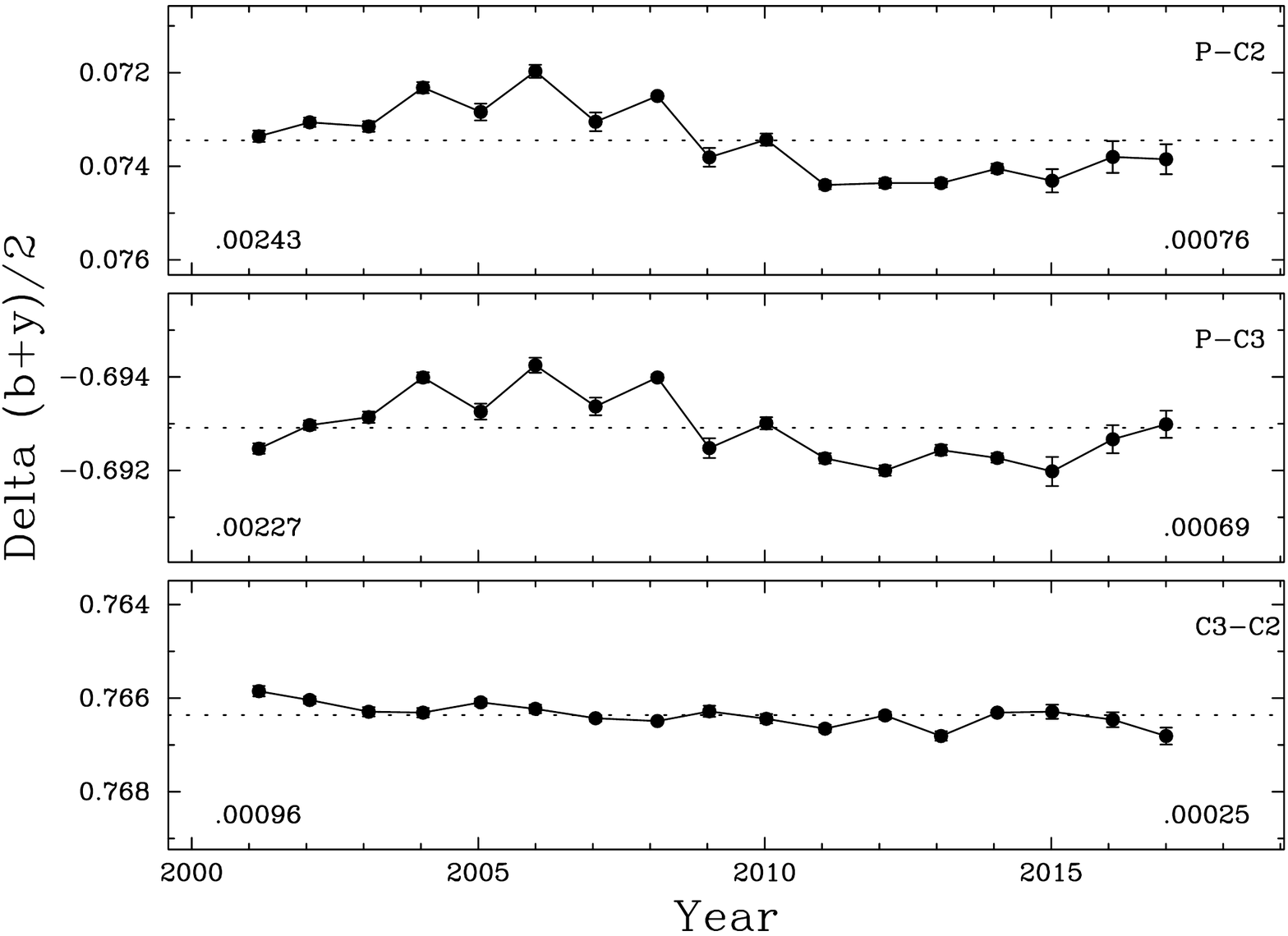}
\caption[]{APT optical photometry of 55\,Cnc. ({\it First and second panels}): Seasonal mean differential magnitudes of 55\,Cnc (P) with respect to comparison stars C2 and C3. ({\it Third panel}): Seasonal mean differential magnitudes between comparison stars C3 and C2. The dotted lines show the grand means for all three panels. The total range of the observations and the standard deviation of the seasonal means from the grand mean are shown in the lower left and lower right of each panel, respectively. The $C3-C2$ differential magnitudes show excellent stability of $\pm0.00025$~mag, demonstrating that the variability in the $P-C2$ and $P-C3$ light curves is intrinsic to 55~Cnc.}
\label{fig:55cnc_cycle_APT}
\end{figure}
\end{center}

\begin{center}
\begin{figure}[h!]
\centering
\includegraphics[trim=0cm 0cm 0cm 0cm,clip=true,width=\columnwidth]{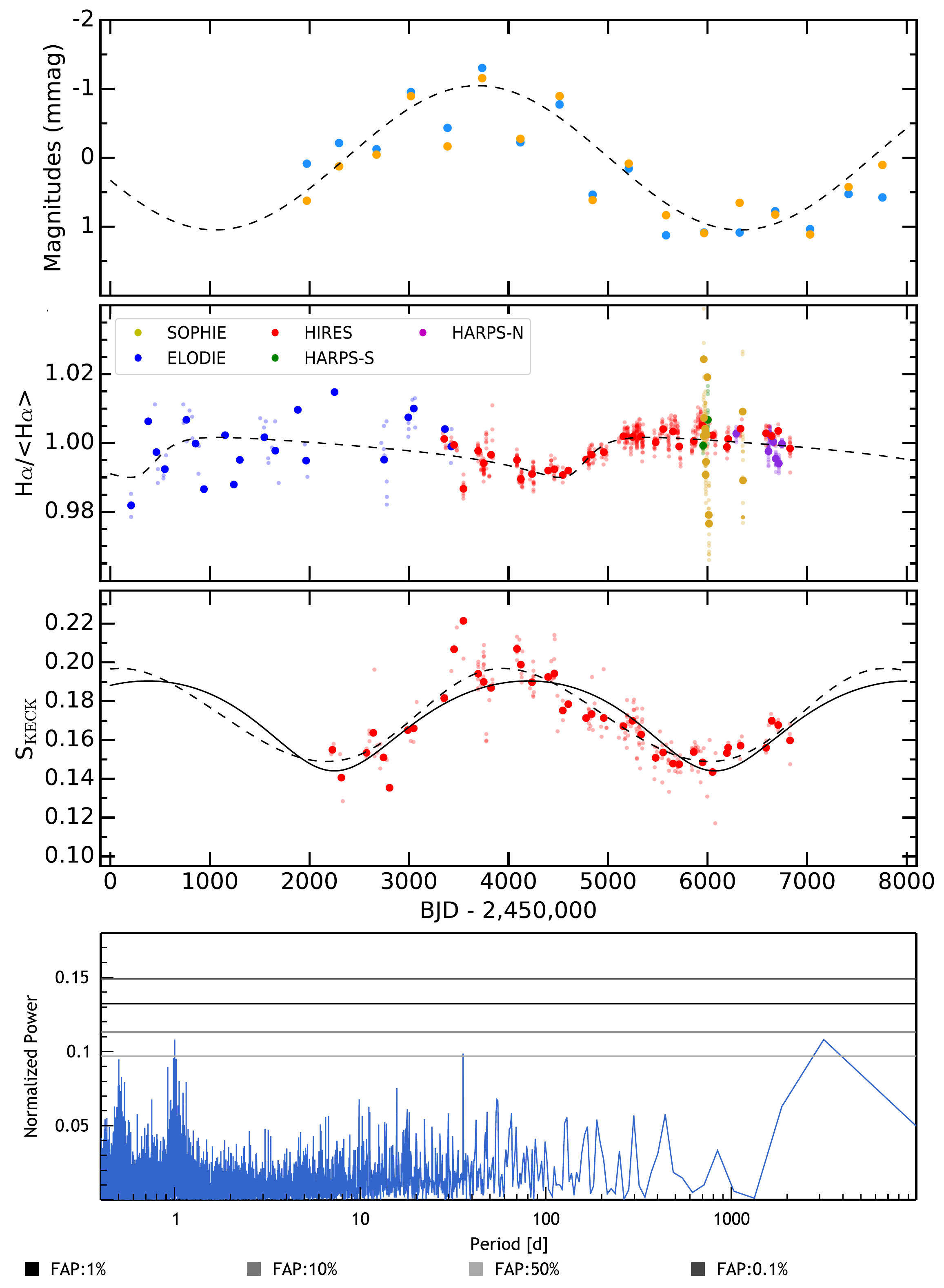}
\caption[]{Magnetic cycle of 55\,Cnc. \textit{First panel:} Seasonal mean differential magnitudes of 55\,Cnc with comparison stars C2 (blue) and C3 (orange), with their best-fit sine function shown as a dashed black curve. \textit{Second panel:} 55\,Cnc H$\alpha$ activity index and the best-fitted Keplerian to the data (dashed black curve). \textit{Third panel:} 55\,Cnc S activity index derived from KECK HIRES data. The continuous curve represents the best fitted Keplerian to those data, while the dashed curve corresponds to the best Keplerian fitted to the RVs to account for the RV effect of the stellar magnetic cycle (see Table \ref{55CANCRI_tab-mcmc-Summary_params} in Sect. \ref{sec:vel_ana}). \textit{Fourth panel:} Periodogram of the S-index residuals after removing the dashed curve seen in the third panel. The good match between the dashed curve and the continuous one in the third panel, as well as the absence of significant peaks in the periodogram of the S-index residuals, tells us that the extra Keplerian fitted to the RVs in Sect.~\ref{sec:vel_ana} accounts well for the magnetic cycle effect.}
\label{fig:mag_cycle}
\end{figure}
\end{center}

The first several years of our observations reveal all three comparison stars to be constant to the limit of our nightly precision ($\sim$1\,milli-mags). We also find the seasonal means of comp stars $C3$ and $C2$ to be constant to the limit of our yearly precision ($\sim$0.2\,milli-mags). However, comp $C1$ exhibited year-to-year variability over a range of $\sim$3\,milli-mags or more. Therefore, we concentrated our photometric analyses on the differential magnitudes $P-C3_{by}$, $P-C2_{by}$, and $C3-C2_{by}$, whose annual means are given in Table~\ref{tab:photom} and plotted in Fig.~\ref{fig:55cnc_cycle_APT}. The differential magnitudes of 55\,Cnc with comp stars $C3$ and $C2$ revealed similar variations, with a peak-to-peak amplitude of $\sim$2\,milli-mag, significantly larger than the variability in $C3-C2_{by}$. A sine curve fitted to $P-C3_{by}$ and $P-C2_{by}$ (Fig.~\ref{fig:mag_cycle}) yields a period of about 14.4\,years for this periodic variation, which we attribute to the magnetic cycle of 55\,Cnc. \\

 We further monitored 55\,Cnc from the ground, with five different spectrographs, yielding measurements of the H$\alpha$ and S activity indexes over $\sim$20\,yr and $\sim$13\,yrs, respectively (see second and third panel in Fig.~\ref{fig:mag_cycle}). Although the older observations with ELODIE are not extremely constraining, both indexes clearly show periodic variations arising from the magnetic cycle of 55\,Cnc. The S-index data obtained by KECK HIRES highlights a solar-like cycle, with an amplitude of 0.024 \citep[nearly twice that of the Sun which goes from S=0.165 to 0.18, e.g.,][]{Egeland2017} and a similar period, in this case $\sim$10.5$\pm$0.3 years. This result is consistent with the cycle period of 12.6$\stackrel{+2.5}{_{-1.0}}$\,years estimated by \citet{Baluev2015} from the analysis of RV data. We further derive a consistent period of $\sim$11.8\,yrs from the H$\alpha$ index; however, we note that the cycle has opposite phase in the S-index. Such an anti-correlation between H$\alpha$ and S-index have already been observed in other stars, however without a clear explanation \citep[][]{GomesdaSilva2014}. We see a good correlation between the brightness variation in photometry and the S-index, which can be explained by a faculae to spot ratio increasing with activity, similar to what is observed in the Sun \citep[][]{Meunier2010a}. \\

\subsection{Rotation period of 55\,Cnc}
\label{sec:Prot}

\begin{center}
\begin{figure}[h!]
\centering
\includegraphics[trim=0cm 0cm 0cm 0cm,clip=true,width=\columnwidth]{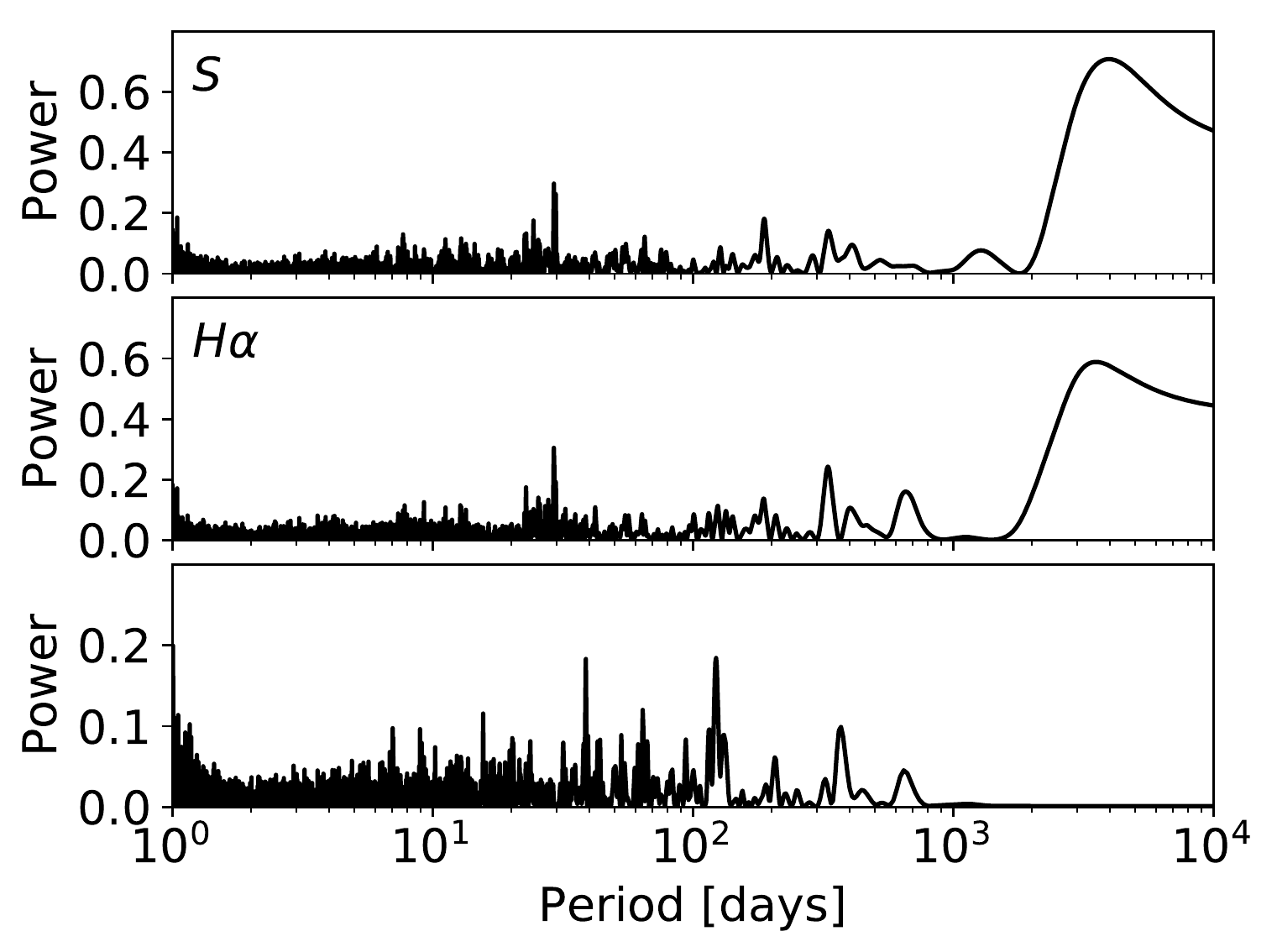}
\caption[]{Periodograms of 55\,Cnc S and H$\alpha$ activity indexes, showing power at the period of the stellar magnetic cycle. The bottom panel corresponds to H$\alpha$ after correcting for the cycle.}
\label{fig:mag_cycle_actind}
\end{figure}
\end{center}

\begin{center}
\begin{figure}[h!]
\centering
\includegraphics[trim=0cm 0cm 0cm 0cm,clip=true,width=\columnwidth]{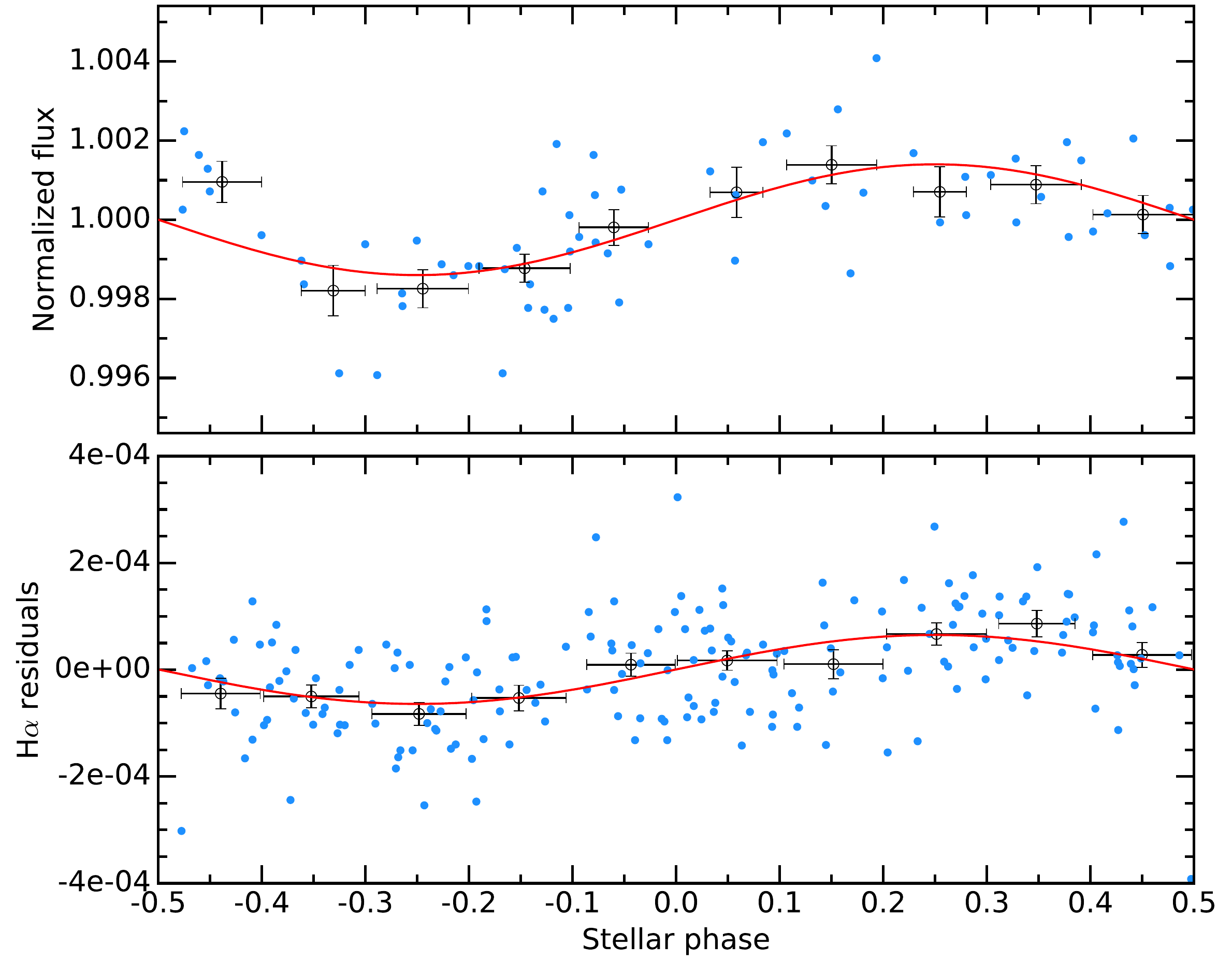}
\caption[]{Rotational modulation of 55\,Cnc in season 9 optical photometry (\textit{upper panel}, $P_\mathrm{*}$ = 40.3\,days) and in the H$\alpha$ index (\textit{lower panel}, $P_\mathrm{*}$ = 38.8\,days). Data has been corrected for the stellar magnetic cycle and for the transit of 55\,Cnc\,e (upper panel), then phase-folded to the periods detected in each dataset individually. Binned data points (black points) allow for a better comparison with the best-fit sine variations (red line).}
\label{fig:55Cnc_rot_modul} 
\end{figure}
\end{center}

After correcting for the stellar magnetic cycle, we identified a sharp peak in the periodogram of H$\alpha$ residuals at about 39\,days (Fig.~\ref{fig:mag_cycle_actind}). A fit to the H$\alpha$ residuals using a sine curve then yielded a period of 38.8$\pm$0.05\,days. The difference in Bayesian information criterion (BIC) between this model and a constant value is about 30, confirming that we detect a significant signal in the activity index of 55\,Cnc, which we associate with the stellar rotational modulation (Fig.~\ref{fig:55Cnc_rot_modul}). \\

Differential magnitudes of 55\,Cnc were corrected for long-term variations and normalized so that each observing season has the same mean. We first performed a frequency analysis based on least-squares sine fits and obtained a clear detection at 40.4$\pm$0.8\,days in season 9, which we attribute to rotational modulation in the visibility of surface starspots (Fig.~\ref{fig:55Cnc_rot_modul}). We then compared the difference in BIC in each season between best-fits of a constant value and of a sine function initialized at the detected period. The error on the measurements was set to their overall standard deviation. In our models we included the best-fit transit function derived in Sect.~\ref{sec:tr_APT_ana} to avoid information loss. Sine functions yielded comparable or improved BIC values in seasons 5 to 10, and 17, which correspond to higher level of activity in the S-index. During a maximum activity phase, it is expected that more spots will be present on the solar surface, therefore inducing a stronger photometric variability (see Fig.~\ref{fig:mag_cycle}). Excluding Season 8, which exhibits a significantly larger periodicity at about 65 days, the other seasons show similar periods ranging from $\sim$35 to 46 days (Table~\ref{tab:photom}). Their weighted mean, 40.04$\pm$0.39 days, is consistent with the period measured in Season 9 and in the H$\alpha$ residuals, and we adopted the latter as final best estimate for the rotation period of 55\,Cnc (P$_\mathrm{rot}$ = 38.8$\pm$0.05\,days). This measurement is consistent with the period of $\sim$39\,days derived by \citealt{Henry2000} from monitoring of the Ca\,II emission, and from previous estimates based on T8 APT photometry (42.7$\pm$2.5\,days, \citealt{Fischer2008}). \\

We measure an average $log$($R^{'}_{HK}$) of -5.03 derived from KECK HIRES S-index data, which yields a rotation period of 49.8$\pm$4.8\,days from the empirical relations in both \citet{Noyes1984} and \citet{Mamajek2008} (we caution that 55\,Cnc is at the edge of the sample used by both authors). \citet{lopez2014} obtained similar results from HARPS and HARPS-N spectra, while \citet{Brewer2016} derived $log$($R^{'}_{HK}$) = -4.98 from a subset of the data we have for KECK HIRES, which yields a rotation period of 46.4$\pm$4.8\,days. Our $log$($R^{'}_{HK}$) further yields an age of $\sim$8.6$\pm$1.0\,Gyr from the relations in \citet{Mamajek2008}, compatible with the value of about 8\,Gyr derived by \citet{Brewer2016} from isochrone fitting, but lower than the ages derived via a similar method by \citet{vonbraun2011} (10.2$\pm$2.5\,Gyr) and \citet{Yee2017} (12.6$\stackrel{+2.9}{_{-2.3}}$\,Gyr). While the low $log$($R^{'}_{HK}$) shows that 55\,Cnc is an old and chromospherically inactive star, we note that it shows temporal variations at X-ray and far-UV wavelengths that trace variability in the upper chromosphere and corona (\citealt{Ehrenreich2012}, Bourrier et al. 2018 submitted). \\

Our measurement of the stellar rotation period was further used to set an upper limit on the projected stellar rotational velocity, $V$\,sin\,$i_{\star} <$ (2\,$\pi$\,$R_\mathrm{\star}$)/$P_\mathrm{\star}$ = 1.23$\pm$0.01\,km\,s$^{-1}$. This value is consistent with the velocity derived from stellar line broadening by \citet{Brewer2016} (1.7$\pm$0.5\,km\,s$^{-1}$) and with the upper limit obtained by \citet{lopez2014} from the nondetection of the Rossiter-McLaughlin effect of 55\,Cnc\,e (0.2$\pm$0.5\,km\,s$^{-1}$). It is, however, about 2.3$\sigma$ lower than the previous velocity derived from stellar line broadening by \citet{Valenti2005} (2.4$\pm$0.5\,km\,s$^{-1}$) and than the velocity derived by \citet{Bourrier2014} from the possible detection of the Rossiter-McLaughlin effect (3.3$\pm$0.9\,km\,s$^{-1}$).


\section{Velocimetric analysis of the 55\,Cnc system}
\label{sec:vel_ana}

The radial velocity (RV) analysis presented here combines all the public data of 55\,Cnc in addition to unpublished out-of-transit SOPHIE data. Therefore, we use the 343 data points from the Tull and HRS spectrograph (\citealt{Endl2012}), the 250 RV measurements from Lick Observatory (\citealt{Fischer2008}), the 629 data points from KECK HIRES (\citealt{Butler2017}), the 292 spectra from HARPS and HARPS-N (\citealt{lopez2014}) and 38 data points from SOPHIE. Those 1552 RV measurements were binned over a timescale of 30 minutes, to average out stellar oscillations and high-frequency granulation (\citealt{Dumusque2011a}). Some observations were taken during the transit of 55\,Cnc\,e, which could affect the RVs due to a Rossiter-McLaughlin effect (\citealt{bourrier2014b}). We note however, that the detection of the Rossiter-McLaughlin effect induced by 55\,Cnc\,e is called into question by \citet{lopez2014}, and even if present, its effect can be neglected in this analysis due to an amplitude smaller than 0.5\ms (\citealt{bourrier2014b}).

To update the orbital parameters of the known planets orbiting 55\,Cnc using all the published data plus the few additional ones taken with SOPHIE, we performed a fit using the tools available on the Data \& Analysis Center for Exoplanet (DACE, available here \url{http://dace.unige.ch}). Besides the five planets already known, we have added an extra Keplerian to take into account the effect of the stellar magnetic cycle. As discussed in \citet{Dumusque2011c} and \citet{Meunier2010b}, such a magnetic cycle can have a correlated counter part in RVs, with an estimated amplitude of $\sim$12\ms \citep[][]{Lovis2011b} considering the stellar properties of 55\,Cnc (B-V=0.87, Teff=5172, feh=0.35, see Table \ref{table:tab_paramsfit}).

All the parameters fitted with their priors can be found in Table \ref{tab-prior}. The planet orbital parameters were initialized by first searching for the planetary signals in a generalized Lomb-Scargle (GLS) periodogram \citep{Zechmeister2009} and then finding the best solution using a Levenberg-Marquardt algorithm. Regarding the Keplerian to account for the effect of the magnetic cycle in the RVs, its initial parameters were set by first fitting the KECK HIRES S-index with a sinusoidal and imposing an amplitude of 12\,\ms, as estimated. Uniform priors were set for all the parameters, except the stellar mass for which we adopted a Gaussian prior  $M_{55\,Cnc} = \mathcal{N}(0.905,0.015)\,M_{\odot}$ from \citet{vonbraun2011} and the period and transit time of 55\,Cnc\,e, for which we also adopted Gaussian priors derived from Spitzer observations $P_{55\,Cnc\,e} = \mathcal{N}(0.73654627800,0.0000018477)$ days and $T_{transit,\,55\,Cnc\,e} =  \mathcal{N}(55733.0058594,1.4648\,10^{-3})$ days (B. Demory priv. comm.). Starting from those initial conditions and priors, we ran the MCMC algorithm available on DACE with 2.10$^6$ iterations.

\begin{center}
\begin{figure*}[h!]
\includegraphics[trim=0cm 0cm 0cm 0cm,clip=true,width=\columnwidth]{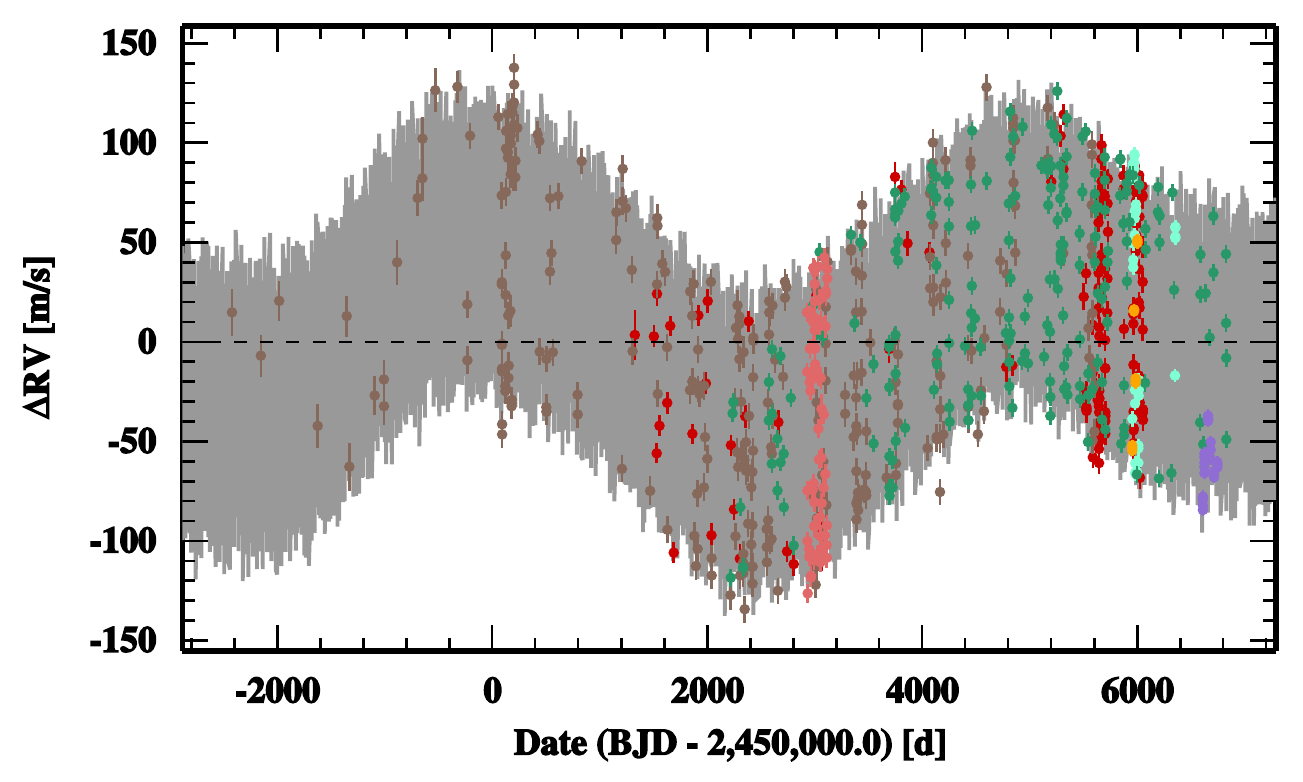}
\includegraphics[trim=0cm 0cm 0cm 0cm,clip=true,width=\columnwidth]{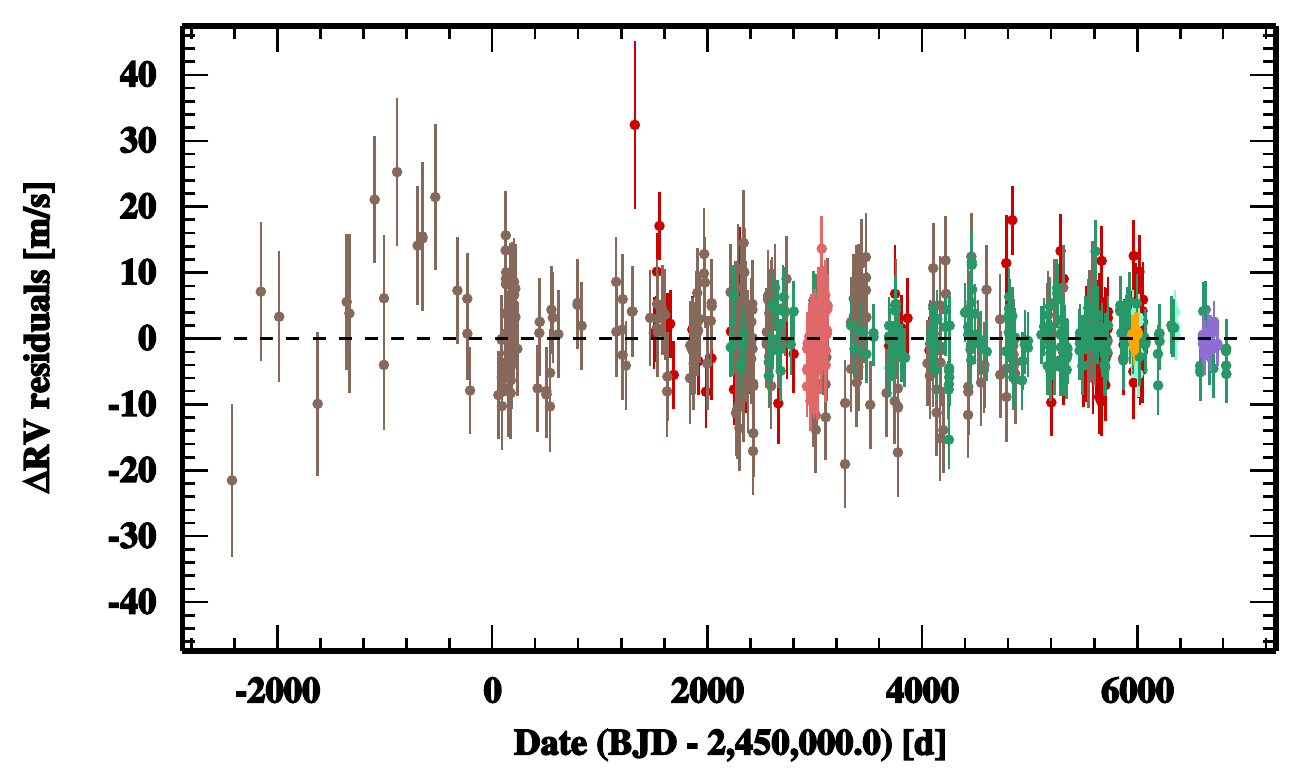}
\includegraphics[trim=0cm 0cm 0cm 0cm,clip=true,width=\columnwidth]{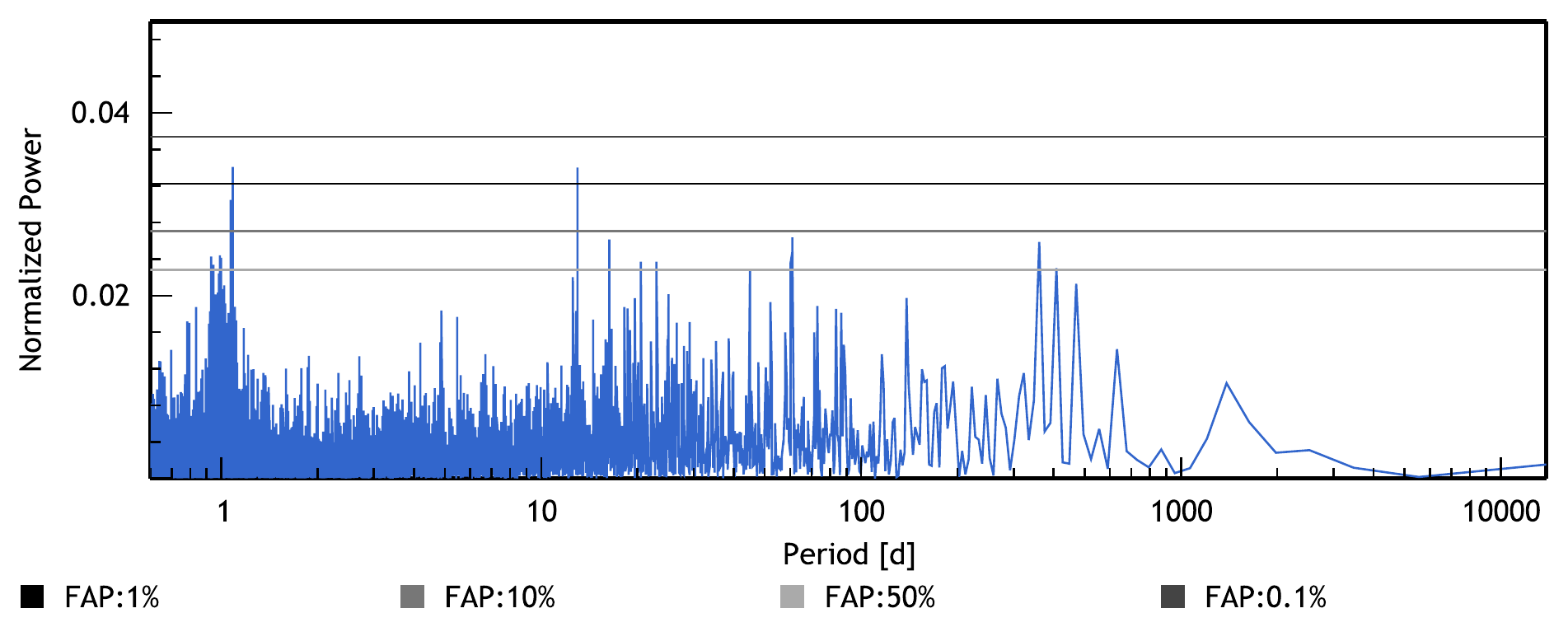}
\includegraphics[trim=0cm -15cm 0cm 0cm,clip=true,width=10cm]{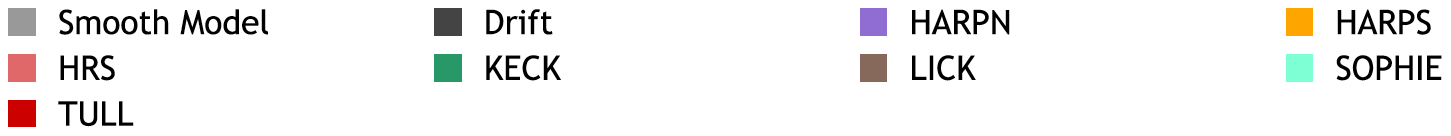}
\caption[]{\emph{Top left:}  The 1552 RV measurements binned over a timescale of 30 minutes with the best model overplotted. This model is obtained by taking the median of each marginalized posterior after our MCMC run. \emph{Top right:} RV residuals after removing the best model. \emph{Bottom left:} GLS periodogram of the RV residuals including false-alarm probability detection thresholds.}
\label{fig:55Cnc_RV_fit_residuals}
\end{figure*}
\end{center}

\begin{center}
\begin{figure*}[h!]
\includegraphics[trim=0cm 0cm 0cm 0cm,clip=true,width=18cm]{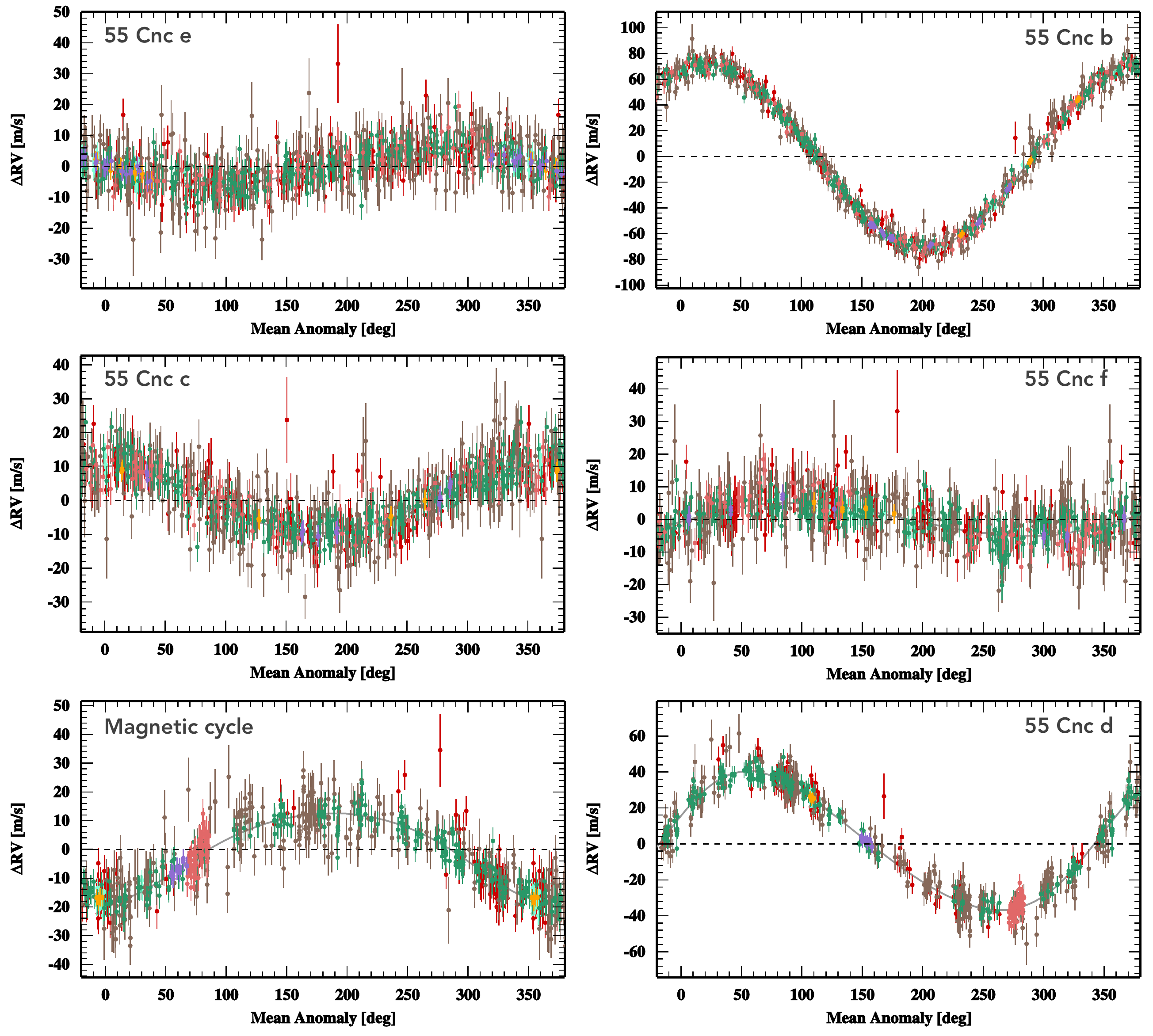}
\includegraphics[trim=0cm 0cm 0cm 0cm,clip=true,width=14cm]{legend_RV_data.png}
\caption[]{RV of each planet and the magnetic cycle, folded in phase using the best-fitted orbital periods. The signals are ordered with orbital periods, therefore, from left to right and top to bottom, we have planet e, b, c, f, the magnetic cycle, and planet d.}
\label{fig:55Cnc_RV_fit}
\end{figure*}
\end{center}

After removing the first 5.10$^5$ iterations to reject the burn-in period and applying a thining of 384 to remove all correlation within the chains, the result of the MCMC can be found in Table\,\ref{tab:55CANCRI_tab-mcmc-Probed_params} of the Appendix. The median of the posterior of the parameters of interest with 68.3\% confidence intervals are shown in Table~\ref{55CANCRI_tab-mcmc-Summary_params}. The best fit using the median of all the marginalized posteriors, as well as the RV residuals and corresponding GSL periodogram is presented in Fig. \ref{fig:55Cnc_RV_fit_residuals}. As we can see, only one signal at 12.9 days has a $p$-value smaller than 1\%. The significance of this signal is not high enough to consider a potential extra planet in the system, and with planet b at 14.7 days, this seems unlikely. This signal is very likely an harmonic of the stellar rotation period, as it corresponds to a third of its value (38.8\,days, Sect.~\ref{sec:Prot}). The RV of each planet and the magnetic cycle folded in phase are shown in Fig. \ref{fig:55Cnc_RV_fit}.

We note that the planetary parameters that we derive are close to the one published in \citet{Endl2012} and \citet{Fischer2017}, however not always compatible within 3\,$\sigma$. The most significant difference comes from planet d, the outermost planet in the system that we found at a period of 5574$\stackrel{+94}{_{-89}}$ days, while it was estimated at 4909$\pm$30 and 5285$\pm$4.5\,days in \citet{Endl2012} and \citet{Fischer2017}, respectively. This large difference can be explained by the fact that we take into account the non-negligible 15 \ms effect of the magnetic cycle in our analysis, which was not done in the past. To make sure that the Keplerian we fitted to account for the magnetic cycle indeed take into account this effect and does not fit any spurious signal in the RV residuals, we show the fitted signal on top of the KECK HIRES S-index in Fig.~\ref{fig:mag_cycle} (third panel, dashed line). We only adjusted the amplitude as RV and S-index are not on the same scale. The strong correlation between this Keplerian and the S-index variation and no significant signal in the S-index residuals after removing this Keplerian (Fig.~\ref{fig:mag_cycle} fifth panel) tell us that this extra component takes correctly into account the RV effect induced by the magnetic cycle of 55\,Cnc. Other significant differences can be seen for the eccentricity of planet e and f, \citet{Fischer2017} reporting significant eccentricities of 0.22$\pm$0.05 and 0.27$\pm$0.05, respectively. Our solution converges to smaller eccentricities of 0.05$\pm$0.03 and 0.08$\pm$0.05, compatible within 0 at 2\,$\sigma$. The smaller eccentricity for planet e is in agreement with orbital circularization, which is expected for such short orbital periods.

\begin{table*}
\caption{List of parameters probed by the MCMC. The symbols $\mathcal{U}$ and $\mathcal{N}$ used for the priors definition stands for uniform and normal distributions, respectively.
}
\label{tab-prior}
\tiny
\begin{tabular}{lcclc}
\hline
\hline
Parameters & Units & Priors & Description\\
\hline
 & & & &\\
\multicolumn{5}{l}{\bf Parameters probed by MCMC}\\
 & & & &\\
M$_{\star}$ & [M$_{\odot}$]  & $\mathcal{N}$(0.915, 0.015) & Stellar mass of 55\,Cnc\\
$\sigma_{JIT}$ & [\ms] & $\mathcal{U}$ & Stellar jitter\\
$\sigma_{(HRS,\,Tull,\,Lick,\,KECK,\,HARPN,\,HARPS,\,SOPHIE)}$ & [\ms] & $\mathcal{U}$ & Instrumental jitter\\
$\gamma_{(HRS,\,Tull,\,Lick,\,KECK,\,HARPN,\,HARPS,\,SOPHIE)}$& [\kms]&$\mathcal{U}$ & Constant velocity offset\\
$P_{55\,Cnc\,e}$ & [days] & $\mathcal{N}$(0.73654627800, 1.8477\,10$^{-6}$) & Period\\
$\log{(P)}$ (55\,Cnc\,b, c, f, d, magn. cycle) & log([days]) & $\mathcal{U}$ & Logarithm of the period\\
$\log{(K)}$ (55\,Cnc\,e, b, c, f, d, magn. cycle) & log([\ms]) & $\mathcal{U}$ & Logarithm of the RV semi-amplitude\\
$\sqrt{e}\,\cos{\omega}$ (55\,Cnc\,e, b, c, f, d, magn. cycle) & - & $\mathcal{U}$ & \\
$\sqrt{e}\,\sin{\omega}$ (55\,Cnc\,e, b, c, f, d, magn. cycle)  & - & $\mathcal{U}$ & \\
$T_{transit}$ (55\,Cnc\,e) & [d] & $\mathcal{N}$(55733.0058594, 1.4648\,10$^{-3}$) & Transit time for 55\,Cnc\,e\\
$\lambda_{0}$ (55\,Cnc\,b, c, f, d, magn. cycle) & [deg] & $\mathcal{U}$ & Mean longitude\\
 & & & &\\
\multicolumn{5}{l}{{\bf Physical Parameters derived from the MCMC posteriors (not probed)}}\\
 & & & &\\      
 $P$ & [d] & - & Orbital period & \\
 $K$ & [m\,s$^{-1}$] & - & RV semi-amplitude & \\
 $e$ & - & - & Orbital eccentricity & \\
 $\omega$ & [deg] &  - & Argument of periastron & \\
 $T_C$ & [d] &  - & Time of transit or inferior conjunction & \\
 $a$ & [AU] &  - & Semi-major axis of the relative orbit & \\
 M & [M$_{\rm Jup}$] &  - & Mass relative to Jupiter (when the inclination $i$ is known) & \\
 M & [M$_{\rm Earth}$] &  - & Mass relative to Earth (when the inclination $i$ is known) & \\
 M.$\sin{i}$ & [M$_{\rm Jup}$] &  - & Minimum mass relative to Jupiter & \\
 M.$\sin{i}$ & [M$_{\rm Earth}$] &  - & Minimum mass relative to Earth & \\
 \hline
\end{tabular}
\end{table*}
\begin{table*}
\small
\caption{Best-fitted solution for the planetary system orbiting 55\,Cnc. For each parameter, the median of the posterior is considered, with error bars computed from the MCMC marginalized posteriors using a 68.3\% confidence interval. $\sigma_{(O-C)\,X}$ corresponds to the standard deviation of the residuals around this best solutions for instrument X, and $\sigma_{(O-C)\,\mathrm{all}}$ the weighted standard deviation for all the data. All the parameters probed by the MCMC can be found in the Appendix, in Tables~\ref{tab:55CANCRI_tab-mcmc-Probed_params} and~\ref{tab:55CANCRI_tab-mcmc-Probed_params2}}  \label{55CANCRI_tab-mcmc-Summary_params}
\def\arraystretch{1.5}
\begin{center}
\begin{tabular}{lccccccc}
\hline
\hline
Param. & Units & 55\,Cnc\,e & 55\,Cnc\,b & 55\,Cnc\,c & 55\,Cnc\,f & magnetic cycle & 55\,Cnc\,d \\
\hline
$P$ & [d] & 0.73654737$_{-1.44\,10^{-6}}^{+1.30\,10^{-6}}$ & 14.6516$_{-0.0001}^{+0.0001}$  & 44.3989$_{-0.0043}^{+0.0042}$  & 259.88$_{-0.29}^{+0.29}$  & 3822.4$_{-77.4}^{+76.4}$  & 5574.2$_{-88.6}^{+93.8}$  \\
$K$ & [m\,s$^{-1}$] & 6.02$_{-0.23}^{+0.24}$ & 71.37$_{-0.21}^{+0.21}$  & 9.89$_{-0.22}^{+0.22}$  & 5.14$_{-0.25}^{+0.26}$  & 15.2$_{-1.8}^{+1.6}$  & 38.6$_{-1.4}^{+1.3}$  \\
$e$ &   & 0.05$_{-0.03}^{+0.03}$ & 0.00$_{-0.01}^{+0.01}$  & 0.03$_{-0.02}^{+0.02}$  & 0.08$_{-0.04}^{+0.05}$  & 0.17$_{-0.04}^{+0.04}$  & 0.13$_{-0.02}^{+0.02}$  \\
$\omega$ & [deg] & 86.0$_{-33.4}^{+30.7}$ & -21.5$_{-89.8}^{+56.9}$  & 2.4$_{-49.2}^{+43.1}$  & -97.6$_{-51.3}^{+37.0}$  & 174.7$_{-14.1}^{+16.6}$  & -69.1$_{-7.9}^{+9.1}$  \\
$T_C$ & [d] & 55733.0060$_{-0.0014}^{+0.0014}$ & 55495.587$_{-0.016}^{+0.013}$  & 55492.02$_{-0.42}^{+0.34}$  & 55491.5$_{-4.8}^{+4.8}$  & 55336.9$_{-50.6}^{+45.5}$  & 56669.3$_{-76.5}^{+83.6}$  \\
\hline
$a$ & [AU] & 0.0154$_{-0.0001}^{+0.0001}$ & 0.1134$_{-0.0006}^{+0.0006}$  & 0.2373$_{-0.0013}^{+0.0013}$  & 0.7708$_{-0.0044}^{+0.0043}$  & --  & 5.957$_{-0.071}^{+0.074}$  \\
M & [M$_{\rm Jup}$] & 0.0251$_{-0.0010}^{+0.0010}$ & --  & --  & --  & --  & --  \\
M & [M$_{\rm Earth}$] & 7.99$_{-0.33}^{+0.32}$ & --  & --  & --  & -- & --  \\
M.$\sin{i}$ & [M$_{\rm Jup}$] & -- & 0.8036$_{-0.0091}^{+0.0092}$  & 0.1611$_{-0.0040}^{+0.0040}$  & 0.1503$_{-0.0076}^{+0.0076}$  & --  & 3.12$_{-0.10}^{+0.10}$  \\
M.$\sin{i}$ & [M$_{\rm Earth}$] & --& 255.4$_{-2.9}^{+2.9}$  & 51.2$_{-1.3}^{+1.3}$  & 47.8$_{-2.4}^{+2.4}$  & -- & 991.6$_{-33.1}^{+30.7}$  \\
\hline 
$\sigma_{(O-C)\,HARPN}$ & [m\,s$^{-1}$] & \multicolumn{6}{c}{1.20}\\
$\sigma_{(O-C)\,HARPS}$ & [m\,s$^{-1}$] & \multicolumn{6}{c}{0.81}\\
$\sigma_{(O-C)\,HRS}$ & [m\,s$^{-1}$] & \multicolumn{6}{c}{4.38}\\
$\sigma_{(O-C)\,KECK}$ & [m\,s$^{-1}$] & \multicolumn{6}{c}{3.58}\\
$\sigma_{(O-C)\,LICK}$ & [m\,s$^{-1}$] & \multicolumn{6}{c}{6.61}\\
$\sigma_{(O-C)\,SOPHIE}$ & [m\,s$^{-1}$] & \multicolumn{6}{c}{2.02}\\
$\sigma_{(O-C)\,TULL}$ & [m\,s$^{-1}$] & \multicolumn{6}{c}{4.89}\\
$\sigma_{(O-C)\,\mathrm{all}}$ & [m\,s$^{-1}$] & \multicolumn{6}{c}{4.33}\\
$\log{(\rm Post})$ &   & \multicolumn{6}{c}{-2285.8$_{-5.1}^{+4.5}$}\\
\hline
\end{tabular}
\end{center}
\end{table*}

\section{Photometric analysis of 55\,Cnc\,e transit}
\label{sec:photom_ana}

\subsection{Extraction of HST/STIS 1D spectra}
\label{sec:extra_STIS}

Transit observations of 55\,Cnc\,e were obtained with the low-resolution G750L grating of the Space Telescope Imaging Spectrograph (STIS) spectrograph onboard the Hubble Space Telescope (HST) (PI: Benneke, GO program 13665). This grating covers the wavelength range 524 to 1027\,nm with a dispersion of 4.92\,\AA\, per pixel. Three visits were obtained on 30 October 2014 (Visit A$_\mathrm{STIS}$), 10 May 2015 (Visit B$_\mathrm{STIS}$), and 22 May 2015 (Visit C$_\mathrm{STIS}$), each visit consisting in five HST orbits. Observations were taken in ACCUM mode, yielding 28 subexposures the first orbit of all visits, and 34 (Visit A$_\mathrm{STIS}$) or 41 (Visits B$_\mathrm{STIS}$ and C$_\mathrm{STIS}$) subexposures in subsequent orbits. All subexposures have a duration of 36\,s. The last orbit in each visit was taken as a fringe flat intended to help correcting the near-infrared portion ($>$750\,nm) of the spectra from CCD fringing. However, observations were oversaturated and we do not know how that can affect the correction for the fringing effect. To avoid any potential bias in the derivation of 55\,Cnc e transit depth, we thus discarded the region of the G750L spectra affected by fringing redward of 750\,nm. \\

Oversaturation led charges to bleed far along the detector columns. They were retrieved using a custom rectangular extraction aperture that we applied to the flat-fielded science files (FLT) output by the STIS calibration pipeline CALSTIS (eg \citealt{Demory2015}). Some images have a height of 500 pixels along the cross-dispersion axis, instead of 1024 pixels, but we used the same aperture for all images to extract the spectra in a consistent manner. The width of the aperture covers the full length of the dispersion axis (1024 pixels) and its height was set to 390 pixels to retrieve as many charges as possible while allowing for the background spectrum to be measured. The background was averaged within two regions, 40 pixels in height, and starting 8 pixels above and below the edges of the extraction aperture. We removed cosmics and bad pixels from the background spectrum using median filtering with a running window. It was then fitted with a fifth-order polynomial function in each exposure, which was used to correct the extracted stellar spectra. We attributed to the corrected spectra the wavelength tables issued by the CALSTIS pipeline, and aligned the spectra in each visit by cross-correlating them with their overall mean over the visit. We then compared each spectrum with the average of the other spectra in the same HST orbit, identifying pixels with count rates larger than five times the standard deviation. We attributed to these pixels the count rate from the averaged spectrum. This operation was repeated twice, and spectra were carefully checked for any residual spurious features. We removed the bluest 10 pixels in all spectra, as we found they were varying significantly over each visit. All analyses hereafter are performed on the spectra integrated over the range remaining after excluding these pixels (531--750\,nm). \\

\subsection{Analysis of HST/STIS transit light curve}
\label{sec:tr_STIS_ana}

\begin{center}
\begin{figure}[h!]
\centering
\includegraphics[trim=0cm 0cm 0cm 0cm,clip=true,width=\columnwidth]{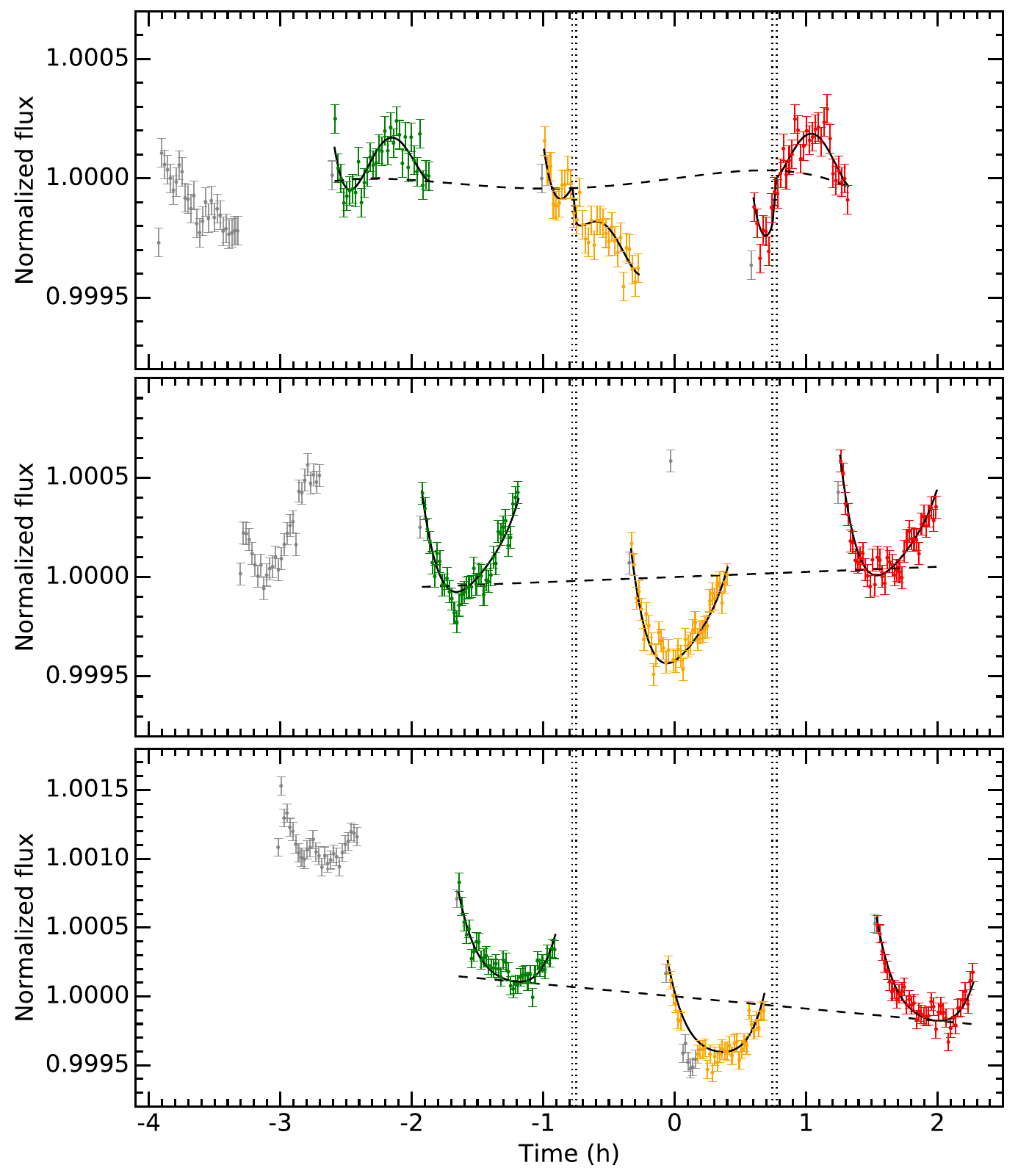}
\caption[]{STIS spectra of 55\,Cnc integrated over the visible band, and plotted as a function of time relative to the transit of 55\,Cnc\,e (vertical dotted lines show the beginning and end of ingress and egress of the transit). The solid black line is the best-fit model to the data, which includes the breathing and long-term flux variations, and the transit. The black dashed line is the model contribution to the long-term variations. Visit A$_\mathrm{STIS}$, B$_\mathrm{STIS}$, and C$_\mathrm{STIS}$ are plotted from top to bottom. The second, third and fourth orbits in each visit are colored in green, orange, and red. Exposures excluded from the fit are plotted in gray.}
\label{fig:LC_STIS_time}
\end{figure}
\end{center}

\begin{center}
\begin{figure}[h!]
\centering
\includegraphics[trim=0cm 0cm 0cm 0cm,clip=true,width=\columnwidth]{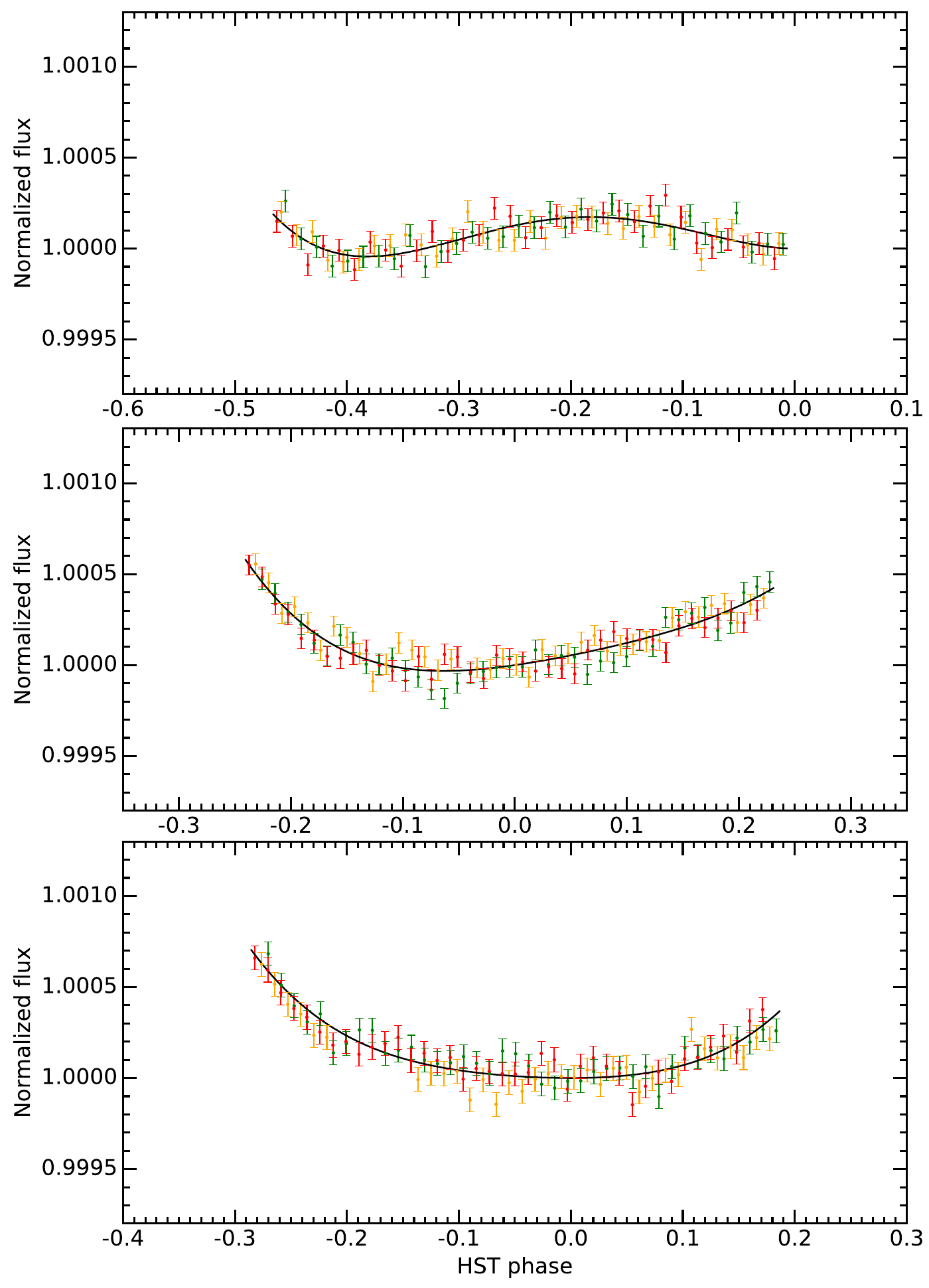}
\caption[]{Visible flux of 55\,Cnc phase-folded on the HST orbital period. Fluxes have been corrected for the long-term variations and transit light curve, to highlight the breathing variations (best-fitted with the solid black line). Color code is the same as in Fig.~\ref{fig:LC_STIS_time}.}
\label{fig:LC_STIS_phase}
\end{figure}
\end{center}

STIS observations are affected by variations in the telescope throughput caused by thermal variations that HST experiences during each orbit (e.g., \citealt{Brown2001}; \citealt{Sing2008a}; \citealt{Evans2013}). This ``breathing'' effect modifies the flux balance within an HST orbit, and is known to be achromatic for a given STIS grating. Our observations display the typical behavior of optical gratings (Fig.~\ref{fig:LC_STIS_time}), with the first orbit showing different flux level and breathing trend than the other orbits. In addition, the first exposure in each orbit shows a significantly lower flux (see eg \citealt{Huitson2012}, \citealt{Sing2013}). First orbits and first exposures were subsequently excluded from our analysis. We also identified long-term variations in each visit that could be linked to instrumental stability or variability in the intrinsic stellar flux (Fig.~\ref{fig:LC_STIS_time}). For all visits we find that a fourth order polynomial function of HST phase was sufficient to describe the breathing variations (Fig.~\ref{fig:LC_STIS_phase}), in agreement with previous studies (eg \citealt{Sing2008a}, \citealt{Demory2015}). The first-order term of this polynomial was found to be unnecessary in Visit C$_\mathrm{STIS}$. Long-term variations are best described with a fourth order polynomial function of time in Visit A$_\mathrm{STIS}$, with the first order term set to zero. In other visits the variations are best described with linear functions. Preliminary fits revealed an outlier in Visit B$_\mathrm{STIS}$ (caused by a spike in the stellar H$\alpha$ line) and six outliers in Visit C$_\mathrm{STIS}$  (likely caused by the transit of a stellar spot), which were excluded from further analysis.  \\

In a second step, we fitted the three visits together using a model that combines the polynomial variations with a transit light curve (calculated with the EXOFAST routines, \citealt{Mandel2002}, \citealt{Eastman2013}). The start and end times of the exposures were converted into BJD$_\mathrm{TDB}$ from the HJD$_\mathrm{UTC}$ times obtained from the file headers (\citealt{Eastman2010}). The model was oversampled in time and averaged within the time window of each exposure before comparison. The parameters of the model are the eight coefficients of the long-term polynomial variations, the ten coefficients of the breathing variations, the planet-to-star radii ratio $R_\mathrm{p}/R_\mathrm{*}$, the mid-transit time at the epoch of our observations $T_\mathrm{0}^\mathrm{STIS}$, the orbital inclination $i_\mathrm{p}$, and the quadratic limb-darkening coefficients $u_\mathrm{1}$ et $u_\mathrm{2}$. We fixed other system properties to the values given in Table~\ref{table:tab_paramsfit}.  \\

\begin{center}
\begin{figure}[h!]
\centering
\includegraphics[trim=1.9cm 0.5cm 0cm 2.7cm,clip=true,width=\columnwidth]{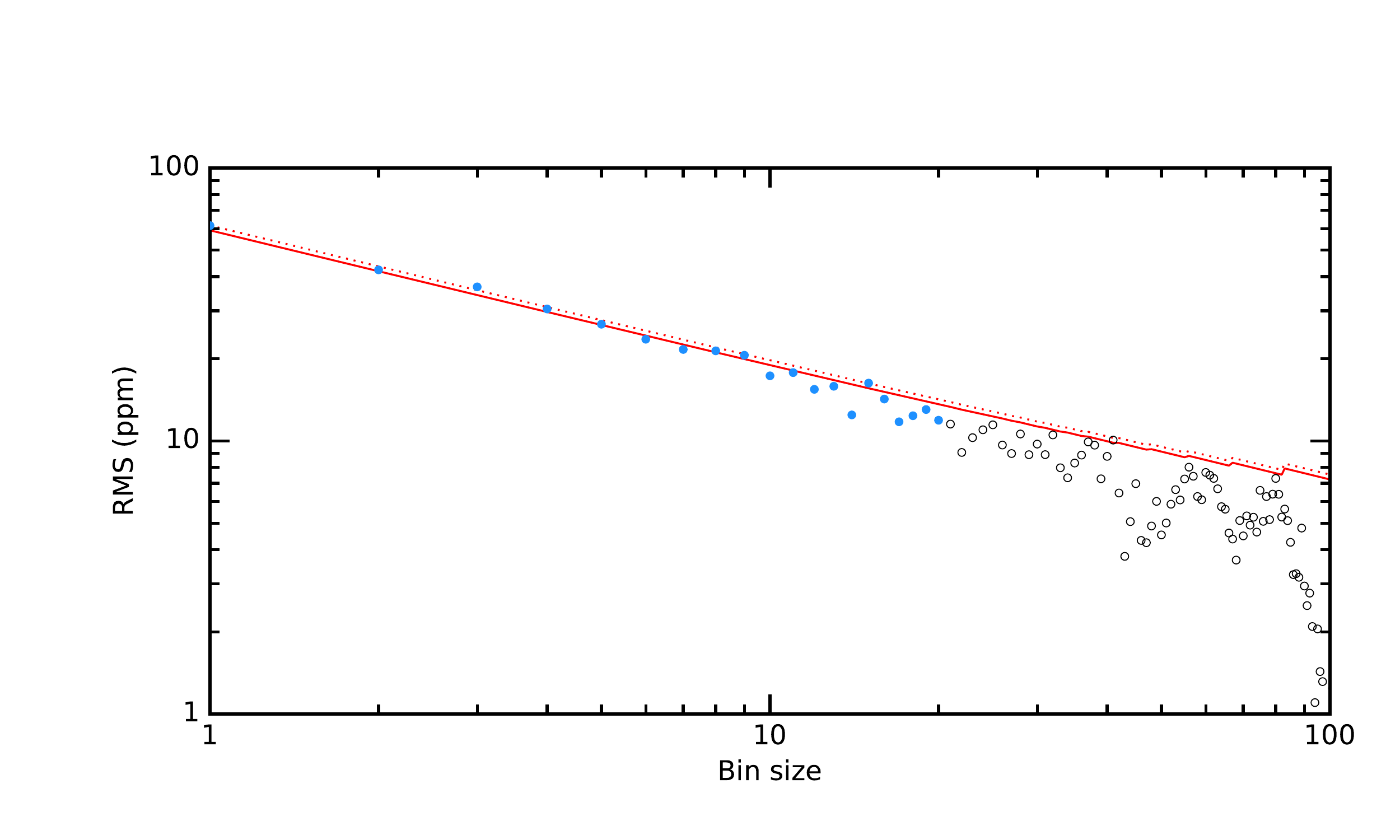}
\caption[]{RMS of binned residuals between the combined STIS visits and their best-fit model, as a function of bin size. The solid red line shows the best-fit noise model (quadratic combination of a Gaussian ``white'' noise and a constant correlated ``red'' noise), which was fitted over the blue measurements. The dotted red line shows a pure Gaussian noise model scaled to the RMS over individual exposures.}
\label{fig:stdev_vs_bin_article}
\end{figure}
\end{center}

\begin{center}
\begin{figure}[h!]
\centering
\includegraphics[trim=0.9cm 0.8cm 1cm 2.7cm,clip=true,width=\columnwidth]{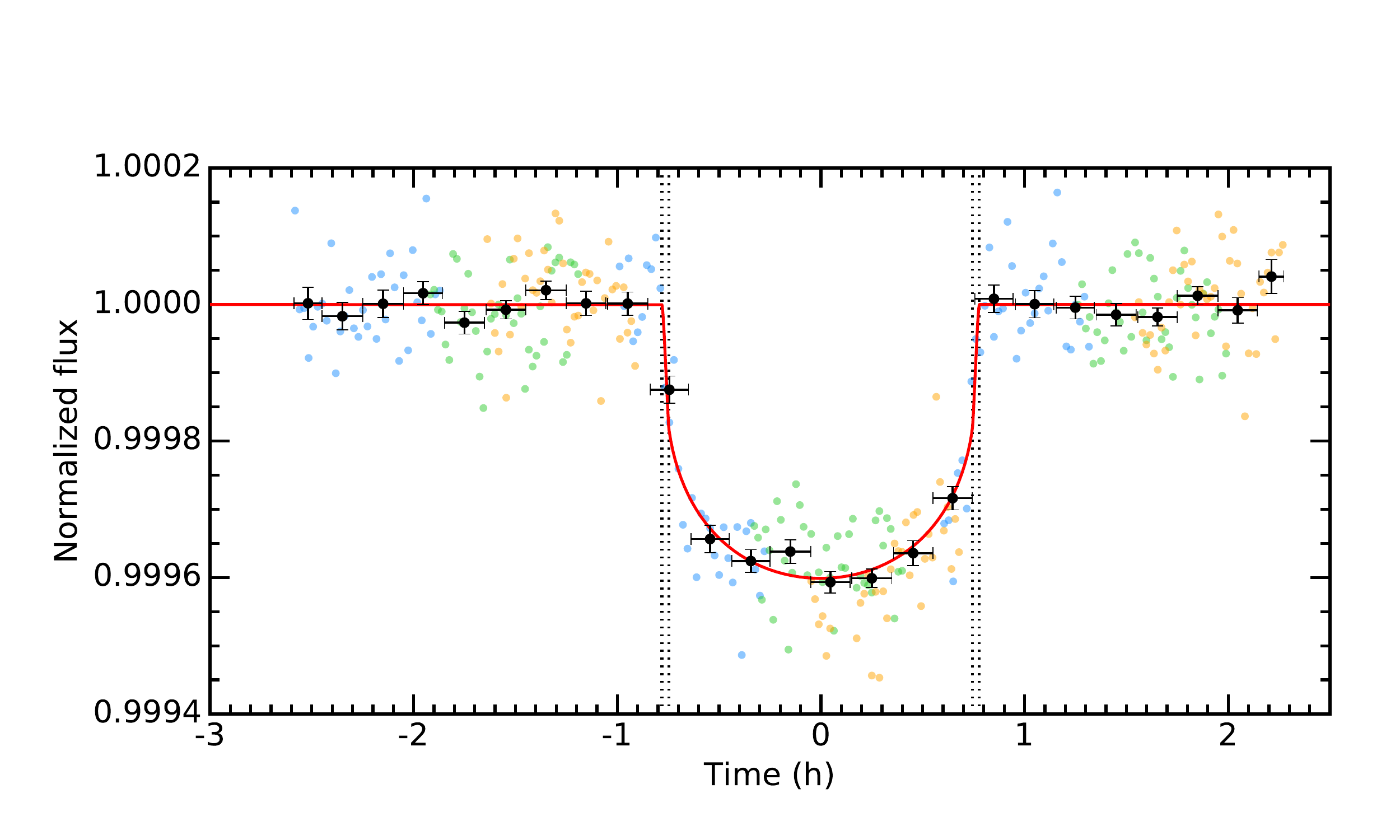}
\caption[]{STIS transit light curve of 55\,Cnc\,e in the visible band. Fluxes have been corrected for the breathing and long-term variations in Visit A$_\mathrm{STIS}$ (blue), B$_\mathrm{STIS}$ (green), and C$_\mathrm{STIS}$ (orange). Black points show binned exposures. The red line is the best-fit transit light curve.}
\label{fig:LC_STIS_corrtime}
\end{figure}
\end{center}

A preliminary fit was obtained with a Levenberg–Marquardt least-squares algorithm. We binned the residuals from this fit within nonoverlapping windows containing $N$ exposures, and measured the standard deviation $\sigma(N)$ of the binned residuals for increasing values of $N$ (e.g., \citealt{pont2006}, \citealt{Winn2009d}, \citealt{Wilson2015}). This revealed that $\sigma(N)$ decreases in $1/sqrt(N)$, which implies that there is no significant correlated noise (see Fig.~\ref{fig:stdev_vs_bin_article}). As a result, we set the uncertainties on the datapoints to the dispersion measured in the residuals (about 60\,ppm in each visit). We then sampled the posterior distributions of the model parameters using the Markov-Chain Monte Carlo (MCMC) Python software package \textit{emcee} (\citealt{Foreman2013}). Model parameters were used as jump parameters, replacing the inclination by its cosine, and the limb-darkening coefficients by the linear combinations $c_\mathrm{1}$ = 2$u_\mathrm{1}$ + $u_\mathrm{2}$ and $c_\mathrm{1}$ = $u_\mathrm{1}$ - 2$u_\mathrm{2}$ (\citealt{Holman2006}). Uniform priors were used with the polynomial coefficients and $R_\mathrm{p}/R_\mathrm{*}$. We imposed Gaussian priors on cos($i_\mathrm{p}$), using the value from \citet{Demory2016b}. We set a Gaussian prior on $T_\mathrm{0}^\mathrm{STIS}$ using the value derived from the velocimetry analysis (Sect.~\ref{sec:vel_ana}), and propagating the uncertainties on $P$ and $T_\mathrm{0}$ onto the time of transit at the epoch of our observations. Quadratic limb-darkening coefficients were estimated in the SDSS r' band (centered at 612.2\,nm with a width of 115\,nm) using the EXOFAST calculator \footnote{\url{http://astroutils.astronomy.ohio-state.edu/exofast/limbdark.shtml}} (\citealt{Eastman2013}) and the stellar temperature, gravity, and metallicity from Table~\ref{table:tab_paramsfit}. We varied these parameters within their 1$\sigma$ uncertainties to determine uncertainties on $u_\mathrm{1}$ and $u_\mathrm{2}$ (found to be dominated by the error on the metallicity). The mean and errors so derived ($u_\mathrm{1}$ = 0.545$\pm$0.008 et $u_\mathrm{2}$ = 0.186$\pm$0.004) were used as Gaussian priors in the MCMC. We initialized 300 walkers that were started at random points in the parameter space, close to the preliminary fit. We ran the walkers for 7000 steps and removed a conservative 3000 steps as burn-in. We checked that all walkers converged to the same solution, before thinning their chains using the maximum correlation length of all parameters. The final thinned and merged chain contains about 4000 independent samples. We set the best-fit values for the model parameters to the medians of the posterior probability distributions and evaluated their 1$\sigma$ uncertainties by taking limits at 34.15\% on either side of the median. Results are given in Table~\ref{table:tab_55Cnce}. The best-fit transit light curve is shown in Fig.~\ref{fig:LC_STIS_corrtime}. Taking into account the uncertainty on the stellar radius, the corresponding planet-to-star radius ratio $R_\mathrm{p}/R_{\star}$ = 0.0182$\pm$0.0002 corresponds to an optical radius $R_\mathrm{p}$ = 1.875$\pm$0.029\,$R_\mathrm{Earth}$. We combined the posterior probability distributions obtained for the mass and radius of the planet to obtain the distribution for the density, and derived $\rho_\mathrm{p}$ = 6.66$_{-0.40}^{+0.43}$\,g\,cm$^{-3}$. \\

\begin{table*}[tbh]
\caption{Final values for the properties of 55\,Cnc\,e}\centering
\begin{threeparttable}
\begin{tabular}{lccl}
\hline
\noalign{\smallskip}  
\textbf{Parameter}      & \textbf{Symbol}                       &           \textbf{Value}                               & \textbf{Unit}          \\   
\noalign{\smallskip}
\hline
\hline
\noalign{\smallskip} 
Planet-to-star radii ratio      & $R_\mathrm{p}/R_{\star}$      &               0.0182$\pm$2$\times$10$^{-4}$                              &   \\
Radius  & $R_\mathrm{p}$        &       1.875$\pm$0.029  & R$_{\rm Earth}$\\
Mass$^{\dagger}$  & $M_\mathrm{p}$   &   7.99$_{-0.33}^{+0.32}$   &  M$_{\rm Earth}$ \\ 
Density  & $\rho_\mathrm{p}$   &   6.66$_{-0.40}^{+0.43}$   &  g\,cm$^{-3}$ \\ 
Transit epoch                            & $T_\mathrm{0}^\mathrm{STIS}$-2~450~000                                 &       7063.2096$\stackrel{+0.0006}{_{-0.0004}}$                & BJD$_{\rm TDB}$  \\
Orbital period$^{\dagger}$                       & $P$                                  &               0.7365474$\stackrel{+1.3\,10^{-6}}{_{-1.4\,10^{-6}}}$    & days  \\
Orbital inclination                             & $i_\mathrm{p}$                                &         83.59$\stackrel{+0.47}{_{-0.44}}$                    & deg   \\
Impact parameter                  & $b$                                 &   0.39$\pm$0.03   &  \\ 
Eccentricity$^{\dagger}$                 & $e$                   &       0.05$\pm$0.03                                                                    &  \\
Argument of periastron$^{\dagger}$       & $\omega$              &   86.0$_{-33.4}^{+30.7}$                  & deg  \\  
Scaled semi-major axis$^{\dagger}$              & $a_\mathrm{p}/R_{\star}$     &     3.52$\pm$0.01  & \\  
Semi-major axis$^{\dagger}$ & $a_\mathrm{p}$  & 0.01544$\pm$0.00005  & au \\  
\hline
\end{tabular}
\begin{tablenotes}[para,flushleft]
Notes: All values are derived from the HST/STIS transit analysis, except for parameters with a dagger derived from the velocimetry analysis, and reported from Table~\ref{55CANCRI_tab-mcmc-Summary_params}. Bulk density is derived from the posterior distributions on the planet mass and radius.
  \end{tablenotes}
  \end{threeparttable}
\label{table:tab_55Cnce}
\end{table*}


\subsection{Analysis of APT transit light curve}
\label{sec:tr_APT_ana}

\begin{center}
\begin{figure}[h!]
\centering
\includegraphics[trim=0.9cm 0.8cm 1cm 2.7cm,clip=true,width=\columnwidth]{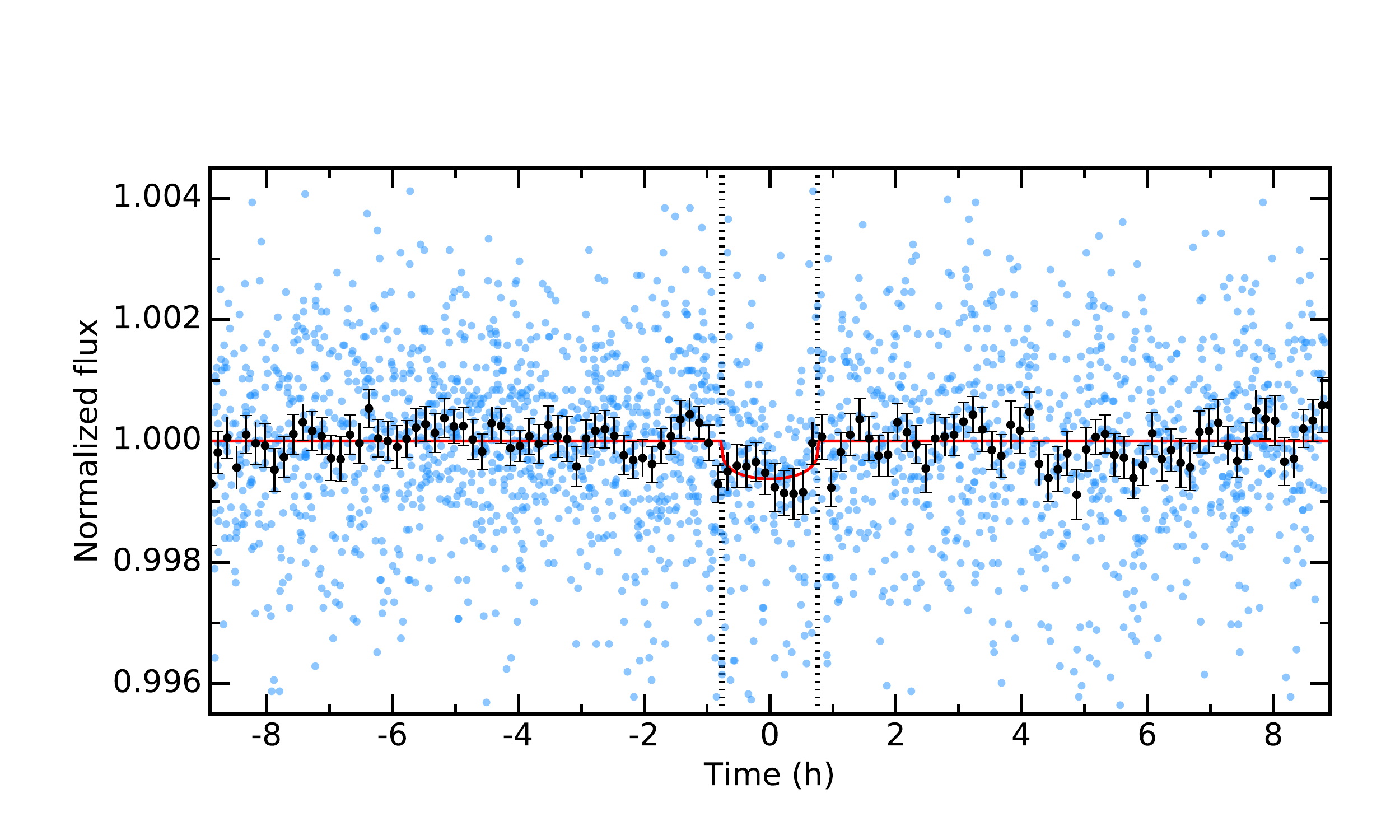}
\caption[]{APT transit light curve of 55\,Cnc\,e in the Str\"omgren b and y bands. Black points show binned exposures. The red line is the best-fit transit light curve.}
\label{fig:LC_APT}
\end{figure}
\end{center}

The transit of 55\,Cnc\,e was detected from the ground by \citet{deMooij2014}, using differential photometry obtained with ALFOSC on the 2.5-m Nordic Optical Telescope. They measured a transit depth of 0.0198$\stackrel{+0.0013}{_{-0.0014}}$ in the Johnson BVR bands, which is consistent with our STIS measurement. We searched for the transit in our normalized APT differential photometry (Sect.~\ref{sec:Prot}) using the EXOFAST model described in Sect.~\ref{sec:tr_STIS_ana}. In a first step, we fitted the transit depth, transit epoch, and orbital period and fixed all other properties to the values given in Table~\ref{table:tab_paramsfit} and Table~\ref{table:tab_55Cnce}. The average Str\"omgren b and y passbands (centered at 467 and 547\AA\,, respectively) overlap with the STIS spectral range, and we consider it reasonable to use the limb-darkening parameters derived in Sect.~\ref{sec:tr_STIS_ana} given the precision of the APT data. Errors on datapoint were set to the dispersion of the residuals from a preliminary best-fit. We found the transit at a period $P$ = 0.736547$\pm$2$\times$10$^{-6}$\,days and epoch $T_\mathrm{0}^\mathrm{APT}$ = 2\,457063.201$\pm$0.007\,BJD$_\mathrm{TDB}$, in good agreement with the results from space-borne photometry (Table~\ref{table:tab_55Cnce}). In a second step we thus fitted the transit depth alone (Fig.~\ref{fig:LC_APT}), all other properties being fixed to their values in Table~\ref{table:tab_55Cnce}. We obtain $R_\mathrm{p}/R_\mathrm{*}$ = 0.0228$\pm$0.0023, which is marginally larger (2$\sigma$) than the STIS value derived in Sect.~\ref{sec:tr_STIS_ana}.\\

\section{Interior characterization of 55\,Cnc\,e}
\label{sec:intern_str}

Successive measurements of the mass and radius of 55\,Cnc e have been used to constrain its interior composition, ranging from a planet with a high-mean-molecular-weight atmosphere (\citealt{demory2011}) to a planet with no atmosphere and a silicate-rich (\citealt{Winn2011}) or carbon-rich (\citealt{Madhusudhan2012}) interior. Our new estimates of planetary radius and mass (Table~\ref{table:tab_55Cnce}) are consistent with previous measurements by \citet{Nelson2014} and \citet{Demory2016b} ($R_\mathrm{p}$ = 1.91$\pm$0.08; $M_\mathrm{p}$ = 8.08$\pm$0.31\,M$_{\rm Earth}$), and their improved precision allow us carry further the interior characterization of 55\,Cnc\,e. We used the generalized Bayesian inference analysis of \citet{Dorn2017a} to rigorously quantify interior degeneracy. We investigated two different scenarios: a dry interior that is comprised of gas and rock only, and a wet scenario in which a nongaseous water layer is present underneath the gas layer.

The data that we considered as input to the interior characterization are planet mass, planet radius, irradiation from the host star (i.e., semi-major axes $a = 0.01544$\,au, stellar effective temperature $T_\mathrm{star} = 0.895 T_{\odot}$, and stellar radius $R_\mathrm{star} = 0.943 R_{\odot}$), as well as the stellar abundances. The stellar abundances are used as a proxy for the bulk composition of the planet. We followed the compilation of \citet{Dorn2017}, where the derived median values are $\fesi$ = 1.86 $\pm$ 1.49, $\mgsi$ = 0.93 $\pm$ 0.77, m$_{\rm CaO}$ = 0.013 wt\%, m$_{\rm Al_2O_3}$ = 0.062 wt\%, m$_{\rm Na_2O}$ = 0.024 wt\%. We note that the uncertainties on the abundance constraints are high ($\sim$80 \%).

\subsection{Method}

The compositional and structural interior parameters that we aim to quantify given the data are given in Table~\ref{tableprior}. In the dry scenario, the planetary interior is assumed to be composed of a pure iron core, a silicate mantle comprising major and minor rock-forming oxides Na$_2$O--CaO--FeO--MgO--Al$_2$O$_3$--SiO$_2$, and a gas layer. In the wet scenario, the mantle is topped by a nongaseous water layer, which can be high-pressure ice and/or superionic water given the high temperature of 55\,Cnc e (see also \citealt{Dorn2017}). A gas layer on top of the water layer is still necessary to impose a pressure high enough to prevent evaporation of water. In both the dry and wet scenarios, the gas layer is assumed to be dominated by gas of mean molecular weights that range from hydrogen to heavy compounds (e.g., N$_2$, H$_2$O, CO$_2$, CO, O$_2$, Na$+$, Ca$^\mathrm{2+}$), i.e., 2.3 $ < \mu <$ 50 [g/mol].

The prior distributions of the interior parameters are listed in Table
\ref{tableprior}. The priors were chosen conservatively. The cubic
uniform priors on \rc and \rsolid reflect equal weighing of masses for
both core and mantle. Prior bounds on $\fesima$ and $\mgsima$ are
determined by the host star's photospheric abundance proxies, whenever
abundance constraints are considered. Since iron is distributed between core and mantle, $\fesi$ only sets an upper bound on $\fesima$. For the gas layer the maximum surface pressure ($P_{\rm batm}$) is determined by the maximum gas mass that a super-Earth can accrete and retain \citep{Ginzburg2016}.

The structural model for the interior uses self-consistent
thermodynamics for core, mantle, and water layer.
For the core density profile, we use the equation of state (EoS) fit
of iron in the hcp (hexagonal close-packed) structure provided by
\citet{Bouchet2013} on ab initio molecular dynamics simulations.
For the silicate mantle, we computed equilibrium mineralogy and density
as a function of pressure, temperature, and bulk composition by
minimizing Gibbs free energy \citep{Connolly2009}.  We assumed an
adiabatic temperature profile within core and mantle. 

For the nongaseous water layer, we followed \citet{Vazan2013} using a quotidian equation of state (QEOS) and above  a pressure of 44.3 GPa, we use the tabulated EoS from \citet{Seager2007}. An adiabatic temperature profile is also assumed for the water layer. If a water layer is present, there must be a gas layer on top that imposes a pressure that is at least as high as the vapor pressure of water.

For the gas layer, we used a simplified atmospheric model for a thin, isothermal atmosphere in hydrostatic equilibrium and ideal gas behavior, which is calculated using the scale-height model. The model parameters that parameterize the gas layer and that we aim to constrain are the pressure at the bottom of the gas layer $P_{\rm batm}$, the mean molecular weight $\mu$, and the mean temperature (parameterized by $\alpha$, see below). The thickness of the opaque gas layer $d_{\rm atm}$ is given by

\begin{equation}
d_{\rm atm}= H \ln \frac{P_{\rm batm}}{P_{\rm out}},
\end{equation}
where the amounts of opaque scale-heights $H$ is determined by the ratio of $P_{\rm batm}$ and $P_{\rm out}$. $P_{\rm out}$ is the pressure level at the optical photosphere for a transit geometry that we fix to 20 mbar (Fortney 2007).
The scale-height $H$ is the increase in altitude for which the pressure drops by a factor of $e$ and can be expressed by
\begin{equation}
H = \frac{T_{\rm atm} R^{*}}{g_{\rm batm} \mu },
\end{equation}
where $g_{\rm batm}$ and $T_{\rm atm}$ are gravity at the bottom of the atmosphere and mean atmospheric temperature, respectively. $R^{*}$ is the universal gas constant (8.3144598 J mol$^{-1}$ K$^{-1}$) and $\mu$ the mean molecular weight.
The mass of the atmosphere $m_{\rm atm}$ is directly related to the pressure $P_{\rm batm}$ as

\begin{equation}\label{massgas}
m_{\rm atm}= 4\pi  P_{\rm batm} \frac{r_{\rm batm}^2}{g_{\rm batm}}.
\end{equation}
where $r_{\rm batm}$ is the radius at the bottom of the atmosphere, respectively. The atmosphere's constant temperature is defined as

\begin{equation}
\label{Tequa}
T_{\rm atm}  = \alpha T_{\rm star} \sqrt{\frac{R_{\rm star}}{2 a}},
\end{equation}
where $R_{\rm star}$ and $T_{\rm star}$ are radius and effective temperature of the host star and $a$ the planet semi-major axis. The factor $\alpha$ accounts for possible cooling of the atmosphere and can vary between 0.5 and 1, which is equivalent to observed ranges of albedos among solar system bodies (0.05 for asteroids up to 0.96 for Eris). Significant warming in the thin gas layers is neglected, which can result in an underestimation of gas layer thicknesses with consequences for the predicted interior parameters for water and rock layers.

Here, we calculate globally-averaged interior profiles that do not account for hemispheric variations. However we note that 55\,Cnc\,e shows temperature variations between day- and nightside estimated to be around 1300 K to 3000 K \citep{Demory2016b,Demory2016a}. A plausible explanation is the presence of a molten magma ocean on the dayside that interacts with the gas envelope above \citep{Elkins2012}, while the rock surface on the nightside of the tidally-locked planet could be solidified. Furthermore, the possibility for variable features in the exosphere and thermal emission of 55\,Cnc\,e \citep{RiddenHarper2016,Demory2016b} call for more complex models that would integrate the dynamic interaction between interior and atmosphere as well as the interaction between planet and star. This, however, is beyond the scope of our study. We refer the reader to model II in \citet{Dorn2017a} for more details on both the inference analysis and the structural model. \\

\begin{table*}[ht]
\caption{Interior parameters and corresponding prior ranges.}  
\label{tableprior}
\begin{center}
\begin{tabular}{lllll}
\hline\noalign{\smallskip}
Layer  &    Parameter & Symbol & Prior range & Distribution  \\
\noalign{\smallskip}
\hline\noalign{\smallskip}
Core    &   Radius   &   $r_{\rm core}$         & (0.01  -- 1) $r_{\rm core+mantle}$ &Uniform in $r_{\rm core}^3$\\
Mantle   &   Composition  &$\fesima$           & 0 -- $\fesistar$&Uniform\\
         &            &$\mgsima$         & $\mgsistar$ &Gaussian\\
         & Radius of rocky interior    &  \rsolid  & (0.01 -- 1) $R_p$& Uniform in $r_{\rm core+mantle}^3$\\
Nongaseous water layer  &  Mass fraction    & \Mwater (wet case)           & (0 - 0.9) $M_p$  &Uniform\\
                &                    & \Mwater (dry case)           & 0  & - \\
Gas layer &  Bottom pressure &  $P_{\rm batm}$ & $P_{\rm out}$-$P_{\rm batm, max}$ &   ln-uniform \\
                &   Temperature coefficient       &  $\alpha$       & 0.5-1   & Uniform \\
                &       Mean molecular weight           &   $\mu$          & 2.3 - 50 [g/mol]   & Uniform \\
\hline
\end{tabular} 
\end{center}
\end{table*}

\subsection{Results}

\begin{figure*}
\includegraphics[width=\textwidth]{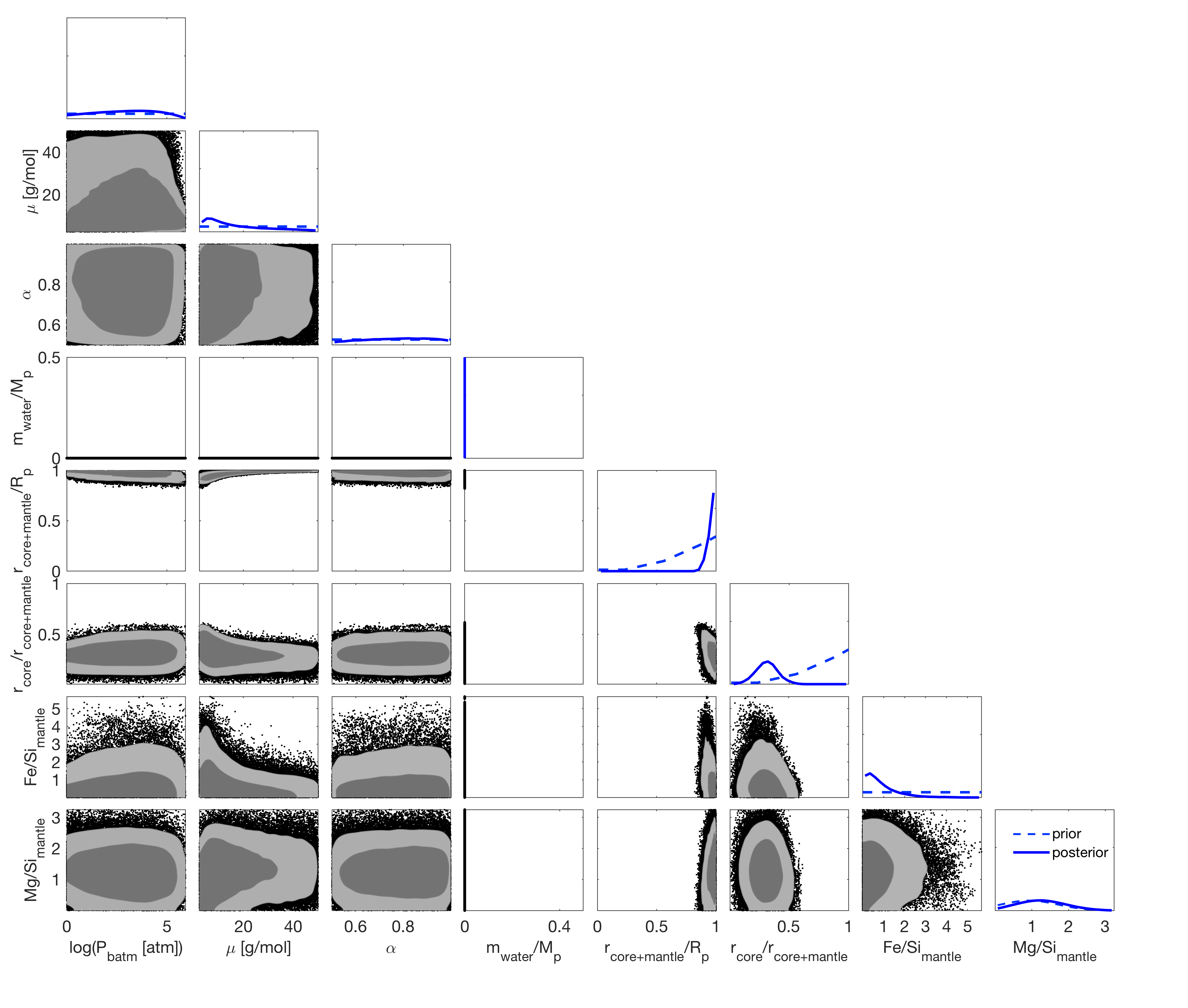}
\caption{Sampled one- and two-dimensional marginal posterior for interior parameters of the dry case: (a) surface pressure $P_{batm}$, (b) mean molecular weight $\mu$, (c) $\alpha$, (d) water mass fraction $m_{\rm water}$/$M_{\rm p}$, which is always zero in the dry case, (e) size of rocky interior $r_{\rm mantle+core}/R_{\rm p}$, (f) relative core size $r_{\rm core}$/$r_{\rm mantle+core}$, (g, h) mantle composition in terms of $\fesima$ and $\mgsima$. The prior distributions are shown in dashed blue.}
\label{Fig:intdry}
\end{figure*}

\begin{figure*}
\includegraphics[width=\textwidth]{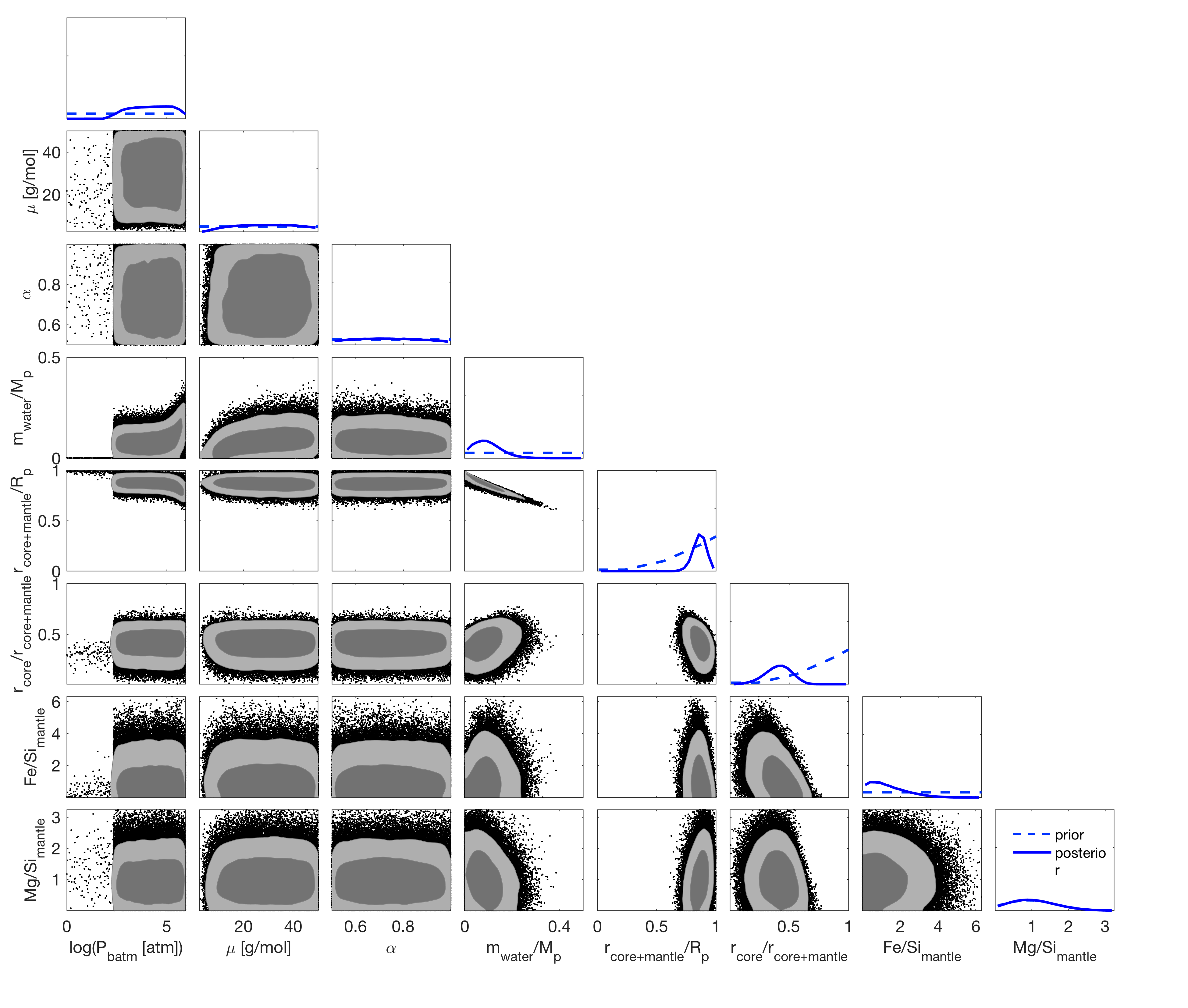}
\caption{Sampled one- and two-dimensional marginal posterior for interior parameters of the wet case: (a) surface pressure $P_{batm}$, (b) mean molecular weight $\mu$, (c) $\alpha$, (d) water mass fraction $m_{\rm water}$/$M_{\rm p}$, (e) size of rocky interior $r_{\rm mantle+core}/R_{\rm p}$, (f) relative core size $r_{\rm core}$/$r_{\rm mantle+core}$, (g, h) mantle composition in terms of $\fesima$ and $\mgsima$. The prior distributions are shown in dashed blue.}
\label{Fig:intwet}
\end{figure*}

Using the generalized McMC method, we obtain a large number of models ($\sim 10^6$) that sample the posterior distribution of possible interiors. For the interior parameters of interest, we obtain posterior distributions that are plotted in Figures \ref{Fig:intdry} and \ref{Fig:intwet}. Parameter estimates are summarized in Table \ref{tableresults}. 

In the dry scenario (Figure \ref{Fig:intdry}) we find interiors that are dominated by a solid interior with radius fractions of \rsolid = $0.97^{+0.02}_{-0.04}$ $R_p$. The corresponding radius of the gas envelope (0.03$^{+0.04}_{-0.02}$\,R$_{\rm p}$ is consistent with an independent analysis of 55\,Cnc e composition performed by \citet{Crida2018}. We note that in the dry scenario the data allows for the complete absence of a gas envelope, but this possibility has to be considered in light of other observations of the planet (Sect.~\ref{sec:scenar_55cnc}). The individual parameters of the gas layer are poorly constrained, except for the thickness of the gas layer.

In the wet scenario, when we allowed for a nongaseous water layer underneath the gas layer, we estimate the possible water mass fraction to be $8_{-4}^{+6}$\%. This result is in agreement with the water mass fraction of $8\pm3$\% estimated by \citet{Lopez2017}. By adding a water layer, we imposed the condition that there must be a gas layer on top that imposes a pressure that is at least as high as the vapor pressure of water. This condition has a major influence on our estimates of gas mas fractions, in other words the gas layer has a minimum surface pressure of 200 atm. In order to fit bulk density while keeping the gas mass high, low  mean molecular weights are excluded. Thick gas and water layers require a smaller rocky interior (\rsolid) in order to fit the total radius. At the same time the total mass can only be fit by a denser rocky interior, which is realized by a larger core size and an iron-enriched mantle while remaining within the bounds of the the abundance constraint $\fesi$. 

The presence of a nongaseous water layer on 55\,Cnc\,e requires a thinner gas envelope (2\% radius fraction) than in the dry scenario (3\% radius fraction). For the wet scenario there is a marginal preference of low-density interiors with denser rocky cores. In order to decide which scenario is more likely, we discuss in Sect.~\ref{sec:scenar_55cnc} our results in light of additional data, specifically those on exosphere observations.

\begin{figure}
\includegraphics[trim=0.cm 0.cm 0.cm 0.cm,clip=true,width=\columnwidth]{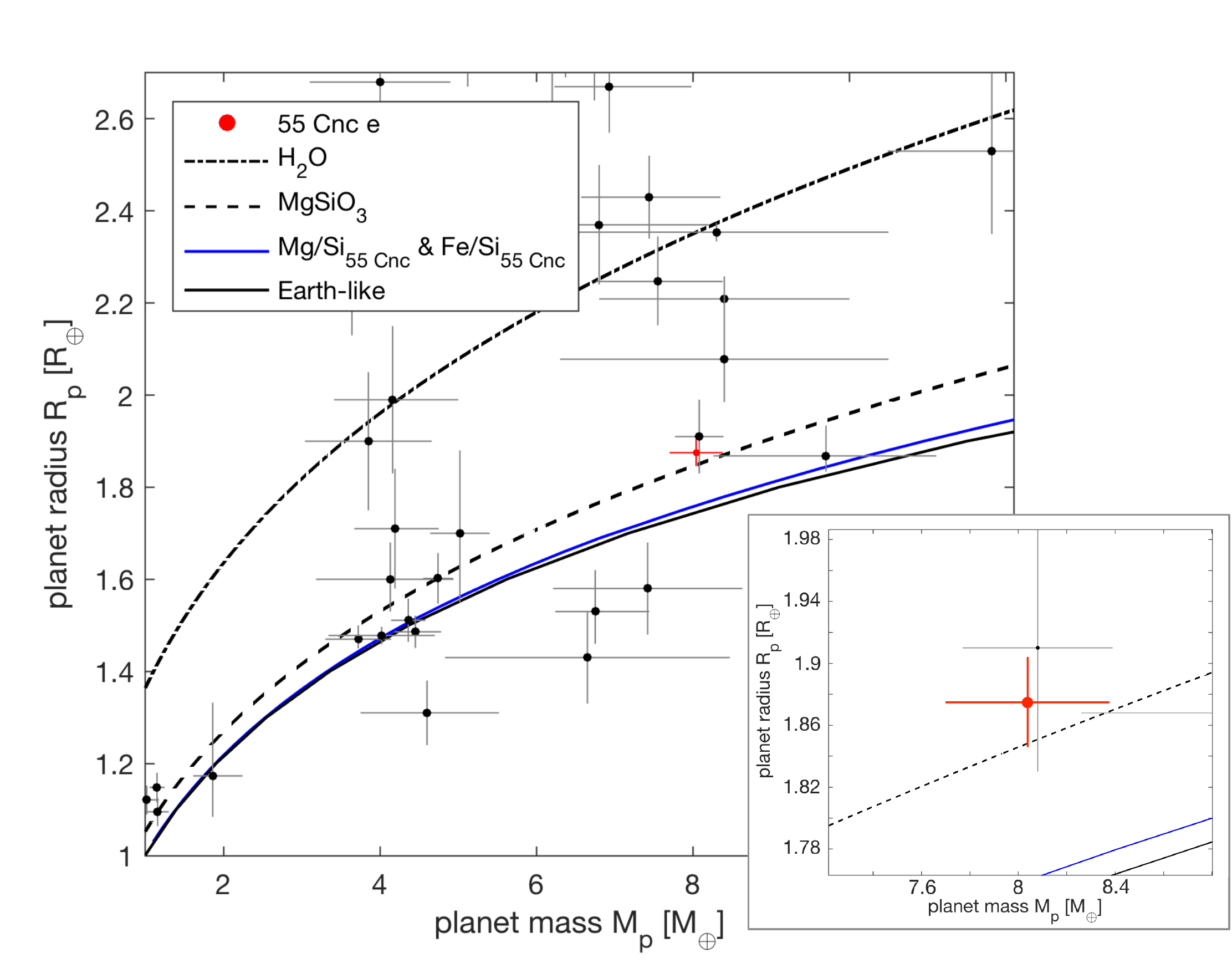}
\caption{Mass and radius of 55 Cnc e (shown in red) in comparison with four mass-radius-relationships of idealized rocky interiors: a pure water composition, the least-dense purely-silicate interior represented by MgSiO$_3$, an interior of an iron core and a iron-free mantle that fits the stellar refractory abundances of 55\,Cnc (mass ratios: Mg/Si = 0.927 and Fe/Si = 1.855), and an Earth-like composition. We show exoplanets with mass known to better than 30\% (error bars represent 1-$\sigma$ uncertainties on their mass and radius).}
\label{Fig:MR}
\end{figure}     

\begin{table}[ht]
\caption{Interior parameter estimates for dry and wet scenario. One-$\sigma$ uncertainties of the one-dimensional marginalized posteriors are listed.} \label{tableresults}
\begin{center}
\begin{tabular}{l|lllll}
\hline\noalign{\smallskip}
Interior parameter & Wet &Dry\\
\noalign{\smallskip}
\hline\noalign{\smallskip}
log$_{10}$({P$_{\rm batm}$ [atm]})  & $4.18_{-1.17}^{+1.11}$ & $2.76_{-2.14}^{+1.77}$ \\
$\mu$ [g/mol]& $28.85_{-13.40}^{+12.98}$ & $13.49_{-7.72}^{+17.85}$ \\
$\alpha$ & $0.74_{-0.15}^{+0.16}$ & $0.78_{-0.16}^{+0.14}$ \\
$r_{\rm gas}$/R$_p$ & $0.02_{-0.01}^{+0.02}$ & $0.03_{-0.02}^{+0.04}$ \\
$m_{\rm water}$/M$_p$ & $0.08_{-0.04}^{+0.06}$ & -- \\
\rsolid /R$_p$& $0.87_{-0.06}^{+0.05}$ & $0.97_{-0.04}^{+0.02}$ \\
$r_{\rm core}$/\rsolid & $0.40_{-0.11}^{+0.10}$ & $0.31_{-0.09}^{+0.09}$ \\
$\fesima$& $1.13_{-0.75}^{+1.22}$ & $0.69_{-0.46}^{+0.90}$ \\
$\mgsima$& $1.01_{-0.55}^{+0.65}$ & $1.25_{-0.62}^{+0.66}$ \\
\hline
\end{tabular} 
\end{center}
\end{table}


\section{Discussion}
\label{sec:disc}

\subsection{Long-term variations in 55\,Cnc\,e radius}
\label{sec:var_temp}

\begin{center}
\begin{figure}[h!]
\centering
\includegraphics[trim=0.9cm 0.8cm 1cm 2.7cm,clip=true,width=\columnwidth]{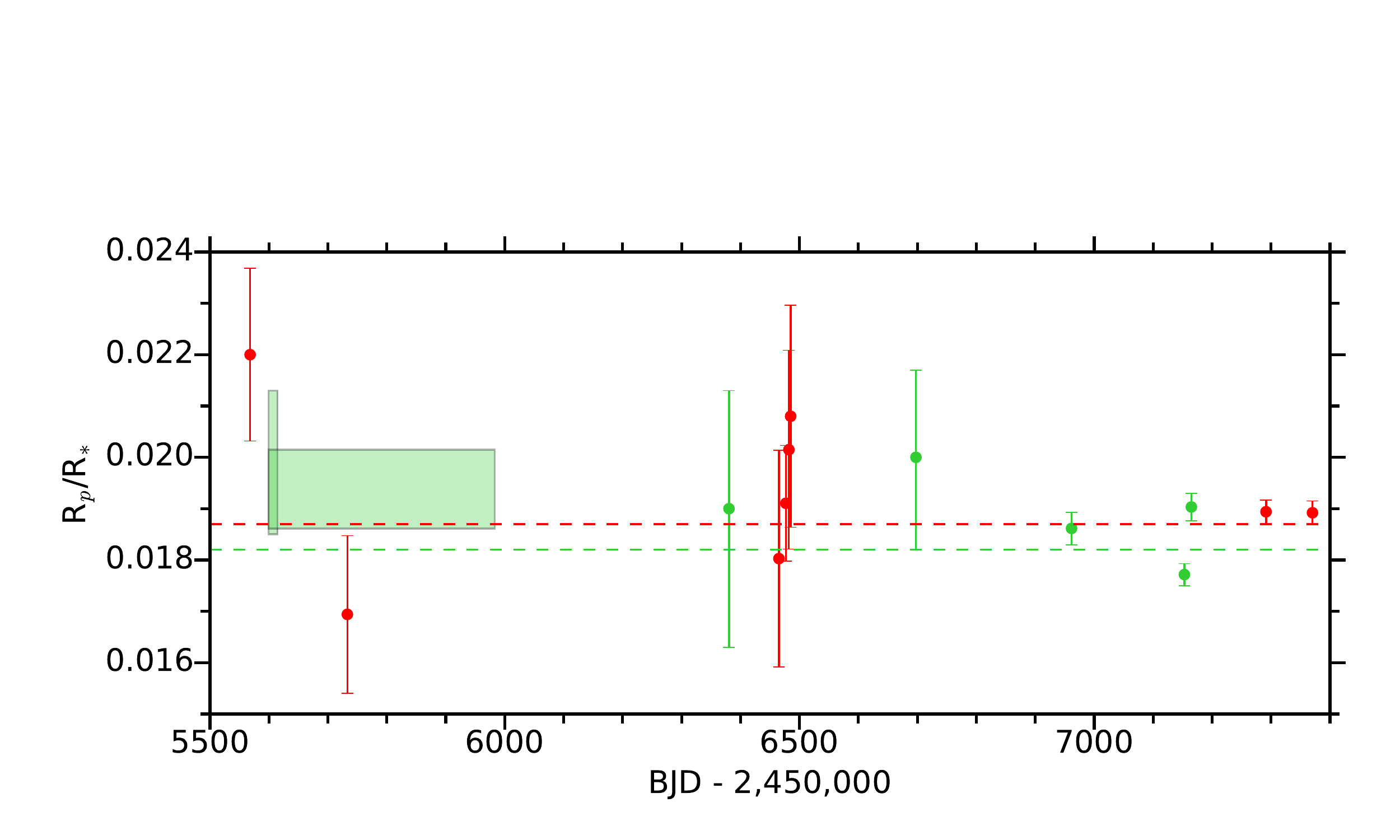}
\caption[]{Measurements of 55\,Cnc\,e planet-to-star radius ratio over time. Green points were obtained in optical bands with MOST (\citealt{Gillon2012} and \citealt{Dragomir2013}), ALFOSC (\citealt{deMooij2014}), HST/STIS (this paper). The first two values are represented as rectangles because they were derived over extended periods of time. The dashed green line shows the value obtained from the fit to the three combined STIS visits. Red points were obtained in infrared bands with Spitzer (\citealt{Demory2016a}) and HST/WFC3 (\citealt{Tsiaras2016}). The dashed red line shows the value obtained from the fit to the combined Spitzer visits (\citealt{Demory2016a,Demory2016b}). }
\label{fig:RpRs_time}
\end{figure}
\end{center}

As mentioned in Sect.~\ref{intro}, there is evidence for a variable source of opacity around 55\,Cnc\,e. This could trace for example temporal variability in an atmosphere subjected to exchange of matter with surface molten rocks as well as losses to space caused by stellar irradiation. This scenario could further be responsible for variations in the apparent radius of 55\,Cnc e over time (\citealt{Demory2016a}). To investigate this possibility, we first fitted the individual transit depth in each HST/STIS visit with the model described in Sect.~\ref{sec:tr_STIS_ana}, all other properties being fixed to their best-fit values (Table~\ref{table:tab_55Cnce}). We derived $R_\mathrm{p}/R_\mathrm{*}$ = 0.0186$\pm$0.0003, 0.0177$\pm$0.0002, and 0.0190$\pm$0.0003 in visits A, B, and C respectively. We show in Fig.~\ref{fig:RpRs_time} those values as a function of time, along with all measurements of 55\,Cnc\,e planet-to-star radius ratio available in the litterature. Ground-based and space-borne transit observations have been obtained between 2011 and 2017 in visible and infrared bands. Apart for marginally lower value in the second of six Spitzer measurements (\citealt{Demory2016a}), early planet-to-star radius ratios of 55\,Cnc\,e measured with MOST (\citealt{Gillon2012}, \citealt{Dragomir2013}), Spitzer (\citealt{demory2011}, \citealt{Gillon2012}, \citealt{Demory2016a,Demory2016b}), and ALFOSC (\citealt{deMooij2014}) are consistent within their uncertainties (Fig.~\ref{fig:RpRs_time}). The five measurements obtained at a much higher precision with HST/STIS (this work) and WFC3 (\citealt{Tsiaras2016}) are consistent with these older values, and with the planet radius we derive from the common fit to the STIS data (within 2.4$\sigma$). There is thus no evidence for long-term variations in the apparent size of 55\,Cnc over timescales of a few years. In contrast, the planet-to-star radius ratio obtained in STIS Visit B is significantly lower ($\sim$4$\sigma$) than the three most recent HST measurements, including the one from Visit C obtained only 12 days later. While we cannot exclude statistical variations, systematic linked to the incompleteness of the STIS individual transits, or stellar variability (although 55\,Cnc is a quiet star at optical wavelengths), the lower planet radius in Visit B might trace temporal variability in 55\,Cnc\,e properties over time-scales of a few days or weeks, as suggested by \citet{Demory2016a}. \\

\subsection{Disentangling between a dry and wet 55\,Cnc\,e}
\label{sec:scenar_55cnc}

\begin{center}
\begin{figure}[h!]
\centering
\includegraphics[trim=0.2cm 0.cm 0.7cm 0.cm,clip=true,width=\columnwidth]{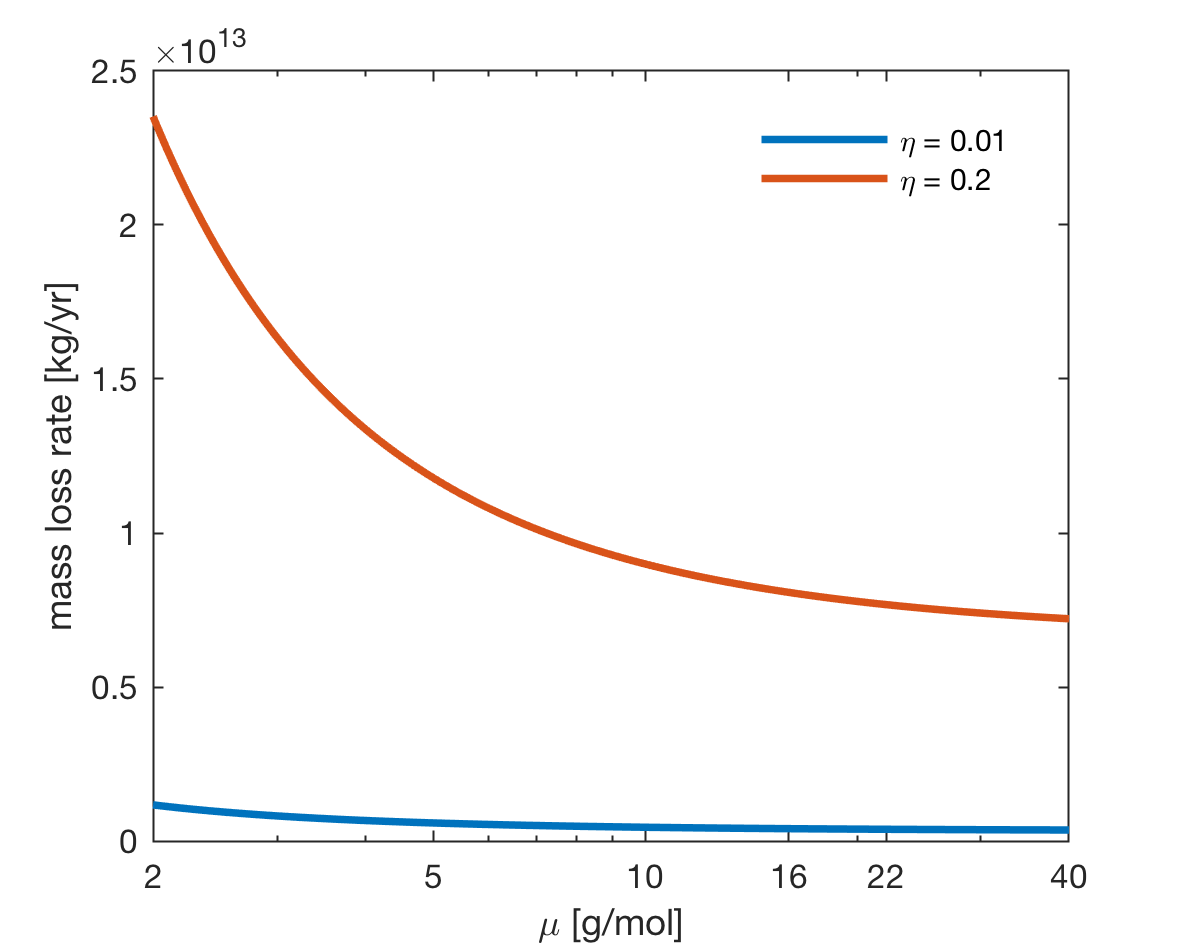}
\caption[]{Mass-loss rate from 55\,Cnc\,e in the energy-limited regime (for a conservative range of efficiencies), as a function of the atmospheric mean molecular weight.}
\label{fig:Fig_MassLossRate}
\end{figure}
\end{center}

For the interior characterization, we have used two scenarios (dry or wet) that differ in their prior assumptions on the presence of a water layer. In either scenarios, we find that a gas fraction likely contributes to the radius by few percents, which can also be inferred from bulk density (Figure \ref{Fig:MR}). However, we find that our interior estimates of the rocky and volatile-rich compounds strongly depend on the scenario. Two implications can be made from this. First, besides the available data, a priori assumptions can contain crucial information on interiors. Second, it is difficult from the measured mass and radius alone to decide whether the wet or the dry case are more likely, which underlines the importance of atmospheric characterization to determine the nature of exoplanets. Our characterization of the interior is solving a static problem and does not account for the evolution of the planet. Over the planet's lifetime, the intense irradiation from the star can lead to significant mass loss from the planetary atmosphere. We plot in Fig.~\ref{fig:Fig_MassLossRate} the atmospheric mass-loss rate from 55\,Cnc e for different mean molecular weights $\mu$, assuming an energy-limited regime (\citealt{Lecav2007}, \citealt{Erkaev2007}) with two representative evaporation efficiencies $\eta$ (0.01 and 0.2, \citealt{Salz2016a}). The mass loss is described as
\begin{equation}
\dot{M} = \frac{\pi \eta F_{\rm XUV} R_{\rm base}^2}{E_g}\,
\end{equation}

where $F_{\rm XUV}$ is the XUV flux at the planet’s age and orbital distance to the star (taken from Bourrier et al. 2018), $E_g$ is the gravitational potential at $R_{\rm base}$. $R_{\rm base}$ is the planet radius at the XUV photosphere
\begin{equation}\label{eq0}
R_{\rm base} \approx R_{\rm p} + H \ln{\left(\frac{P_{\rm photo}}{P_{\rm base}}\right)},
\end{equation}
where $P_{\rm photo}$ and $P_{\rm base}$ are set to typical values of 20 mbar and 1 nbar, respectively \citep{Lopez2017}. The dependency of the mass loss rate on $\mu$ is due to the scale height $H$, which is the scale height in the regime between the optical and the XUV photosphere 
\begin{equation}\label{eq3}
H = \frac{T_{\rm eq} R^{*}}{g_{\rm surf} \mu },
\end{equation}
where $g_{\rm surf}$ is surface gravity and $R^{*}$ is the universal gas constant (8.3144598 J mol$^{-1}$ K$^{-1}$). It is clear that a primordial hydrogen layer would have been lost within a tiny fraction of the planet’s lifetime ($\sim$80\,000\,years), which is consistent with the high mean molecular weights that we derived in the dry ($\mu$ = 13.5$\stackrel{+17.9}{_{-7.7}}$ g/mol) and wet ($\mu$ = 28.9$\stackrel{+13.0}{_{-13.4}}$ g/mol) scenarios. As a natural result of the strongly irradiated water layer in the wet scenario, we would expect the gas layer to be dominated by steam. The planet mass and radius favors a heavier atmosphere in this scenario, and indeed there is no observational evidence for an extended water envelope (\citealt{Esteves2017}). Furthermore, steam would get photodissociated at high-altitudes and sustain an upper atmosphere of hydrogen, which has not been detected (\citealt{Ehrenreich2012}). Finally, Fig.~\ref{fig:Fig_MassLossRate} suggests that this hydrogen envelope would be lost quickly, depleting the steam envelope and underlying water layer. These additional constraints thus favor the dry scenario over the wet scenario.\\

In the dry case we find a gas radius fraction of $r_{\rm gas}$/$R_{\rm p}$ = $3_{-2}^{+4}$\%, which is compatible within 2$\sigma$ with an atmosphere-less body. However, infrared photometric data support the existence of an atmosphere of heavy weight molecules and inefficient heat-redistribution (\citealt{Demory2016b}, \citealt{Angelo2017}). 55\,Cnc e could be surrounded by an atmosphere dominated by rock-forming elements, continually replenished by vaporization of, for example, silicates on top of a possible magma ocean. The existence of such a mineral-rich and water-depleted atmosphere was predicted for hot rocky super-Earths by \citet{Ito2015}, assuming that gas and melted rocks of the magma ocean are in equilibrium. A search for atmospheric escape at FUV wavelengths revealed strong variations in the lines of the star, possibly arising from interactions between 55\,Cnc e and its star, but no clear signature of a metal-rich exosphere (\citealt{Bourrier2018_FUV}). Nonetheless, a mineral atmosphere could explain the inefficient heat-redistribution and relatively high night-side temperature of the planet (\citealt{Demory2016b}, \citealt{Zhang2016}), and would be a likely origin for the sodium and ionic calcium possibly detected in the exosphere of 55\,Cnc e \citep{RiddenHarper2016}. The mean molecular weight of our simple model is consistent with the presence of heavy species such as calcium ($\mu_{\rm Ca} = 40.1$ g/mol), sodium ($\mu_{\rm Na} = 22.99$ g/mol), or oxygen ($\mu_{\rm O} = 16$ g/mol). However, a gas layer dominated by one of these species, and with a radius fraction of 0.03\,$R_{p}$, would have surface pressures of 3700 and 90 bar, respectively (at $\alpha =1$). Such high surface pressures are at odds with the maximum surface pressure of $\sim$0.1 bar estimated by \citet{Ito2015} for a mineral atmosphere in equilibrium with molten rock. The uncertainties we derive for the gas layer thicknesses and mean molecular weights do not allow us to distinguish between a mineral atmosphere and a gas layer dominated by molecules such as CO or N$_2$, as suggested by the infrared phase curve of 55\,Cnc (\citealt{Angelo2017}; \citealt{Hammond2017}).\\


\section{Conclusion}
\label{sec:conc}

We analyzed the long-term activity of 55\,Cnc using two decades of photometry and spectroscopy data. A solar-like cycle ($P_\mathrm{mag}$ = 10.5\,years) is detected in all datasets, along with the stellar rotational modulation ($P_{*}$ = 38.8\,days), confirming that 55\,Cnc is an old ($\sim$10 Gyr) and quiet star. The magnetic cycle was included for the first time in the velocimetric analysis of the system, allowing us to update the orbital and mass properties of the five known planets. Our results are consistent with past publications, except for significant differences in the period of the outermost planet. This is likely because its period is on the same order as that of the magnetic cycle, which we also detect in the radial velocity data. 

The innermost planet 55\,Cnc e is one of the most massive known USP planets, an iconic super-Earth that is a target of choice to understand the formation and evolution of small close-in planets. It orbits one of the brightest exoplanet host-stars, which allowed us to detect its transit in APT ground-based differential photometry. Yet, despite extensive observations across the entire spectrum, the nature of 55\,Cnc\,e remains shrouded in mistery. A precise knowledge of the planet density is necessary to determine its composition and structure, and we combined our derived planet mass ($M_\mathrm{p}$ = 8.0$\pm$0.3\,M$_{\rm Earth}$) with refined measurement of its optical radius derived from HST/STIS observations ($R_\mathrm{p}$ = 1.88$\pm$0.03\,R$_{\rm Earth}$ over 530-750\,nm) to revise the density of 55\,Cnc\,e ($\rho$ = 6.7$\pm$0.4\,g\,cm$^{-3}$). This result suggests that the super-Earth is too light to be purely made of silicate (Fig.~\ref{Fig:MR}), and we thus modeled its interior structure by allowing for water and gas layers. The precision on the mass and radius of 55\,Cnc does not allow us to conclude on the existence of a non-gaseous water layer, but the orbit of the planet at the fringes of the stellar corona ($a_\mathrm{p}$ = 3.5\,$R_{*}$), the nondetection of an extended hydrogen and water atmosphere, the possible detection of exospheric sodium and calcium, and the infrared mapping of the planet, all strongly point toward the absence of a significant water layer. Regarding the gas layer, the bulk density of 55\,Cnc e clearly excludes the presence of a H/He envelope, despite the recent claim by \citet{Tsiaras2016}. Even small amounts of hydrogen would drastically increase the apparent optical radius of the planet, and a hydrogen-rich envelope would not have survived erosion at such close distance from the host star over $\sim$10\,Gy. Instead, we find that a heavyweight atmosphere likely contributes to the planet radius, in agreement with recent results by \citet{Demory2016b}, \citet{Angelo2017}. Degeneracies prevent us from assessing the composition of this envelope, which could include mineral-rich compounds arising from a molten or volcanic surface, or a CO- or N$_2$- dominated atmosphere. In any case the properties of the envelope would have to explain the temporal variability observed in the opacity of the planet. We compared all available measurements of 55\,Cnc\,e planet-to-star radius ratio to search for additional signatures of this variability, and found significant short-term ($\sim$week) variations in the STIS measurements. All measured sizes are nonetheless consistent within 3$\sigma$ with our derived radius, and we found no evidence for long-term variations over timescales of months or years. Observations at high photometric precision and high temporal cadence (e.g., with the CHEOPS satellite), along with high-resolution spectroscopic follow up, will be required to further investigate the variable nature of 55\,Cnc\,e.\\


\begin{acknowledgements}
We thank the referee for their constructive comments. We thank Munazza K. Alam and Mercedes Lopez-Morales for their help with the STIS/G750L observations, and N. Hara and P.A. Wilson for helpful discussions about correlated noise. This work has been carried out in the frame of the National Centre for Competence in Research ``PlanetS'' supported by the Swiss National Science Foundation (SNSF). V.B. acknowledges the financial support of the SNSF. X. Dumusque is grateful to the The Branco Weiss Fellowship – Society in Science for continuous support. C.D. is funded by the Swiss National Science Foundation under the Ambizione grant PZ00P2\_174028. This project has received funding from the European Research Council (ERC) under the European Union's Horizon 2020 research and innovation program (project Four Aces grant agreement No 724427). G.W.H. acknowledges long-term support from NASA, NSF, Tennessee State University, and the State of Tennessee through its Centers of Excellence program. N.A-D. acknowledges support from FONDECYT 3180063. We thank the Programme National de Plan\'etologie for the use of SOPHIE at Observatoire de Haute Provence. 
\end{acknowledgements}

\bibliographystyle{aa} 
\bibliography{biblio} 

\begin{thebibliography}{81}
\expandafter\ifx\csname natexlab\endcsname\relax\def\natexlab#1{#1}\fi

\bibitem[{{Angelo} \& {Hu}(2017)}]{Angelo2017}
{Angelo}, I. \& {Hu}, R. 2017, \aj, 154, 232

\bibitem[{{Baluev}(2015)}]{Baluev2015}
{Baluev}, R.~V. 2015, \mnras, 446, 1493

\bibitem[{{Bouchet} {et~al.}(2013){Bouchet}, {Mazevet}, {Morard}, {Guyot}, \&
  {Musella}}]{Bouchet2013}
{Bouchet}, J., {Mazevet}, S., {Morard}, G., {Guyot}, F., \& {Musella}, R. 2013,
  \prb, 87, 094102

\bibitem[{{Bourrier} {et~al.}(2018){Bourrier}, {Ehrenreich}, {Lecavelier des
  Etangs}, {Louden}, {Wheatley}, {Wyttenbach}, {Vidal-Madjar}, {Lavie}, {Pepe},
  \& {Udry}}]{Bourrier2018_FUV}
{Bourrier}, V., {Ehrenreich}, D., {Lecavelier des Etangs}, A., {et~al.} 2018,
  ArXiv e-prints

\bibitem[{{Bourrier} \& {H{\'e}brard}(2014)}]{bourrier2014b}
{Bourrier}, V. \& {H{\'e}brard}, G. 2014, \aap, 569, A65

\bibitem[{{Bourrier} {et~al.}(2014){Bourrier}, {Lecavelier des Etangs}, \&
  {Vidal-Madjar}}]{Bourrier2014}
{Bourrier}, V., {Lecavelier des Etangs}, A., \& {Vidal-Madjar}, A. 2014, \aap,
  565, A105

\bibitem[{{Brewer} {et~al.}(2016){Brewer}, {Fischer}, {Valenti}, \&
  {Piskunov}}]{Brewer2016}
{Brewer}, J.~M., {Fischer}, D.~A., {Valenti}, J.~A., \& {Piskunov}, N. 2016,
  \apjs, 225, 32

\bibitem[{{Brown} {et~al.}(2001){Brown}, {Charbonneau}, {Gilliland}, {Noyes},
  \& {Burrows}}]{Brown2001}
{Brown}, T.~M., {Charbonneau}, D., {Gilliland}, R.~L., {Noyes}, R.~W., \&
  {Burrows}, A. 2001, \apj, 552, 699

\bibitem[{{Butler} {et~al.}(1997){Butler}, {Marcy}, {Williams}, {Hauser}, \&
  {Shirts}}]{Butler1997}
{Butler}, R.~P., {Marcy}, G.~W., {Williams}, E., {Hauser}, H., \& {Shirts}, P.
  1997, \apjl, 474, L115

\bibitem[{{Butler} {et~al.}(2017){Butler}, {Vogt}, {Laughlin}, {Burt},
  {Rivera}, {Tuomi}, {Teske}, {Arriagada}, {Diaz}, {Holden}, \&
  {Keiser}}]{Butler2017}
{Butler}, R.~P., {Vogt}, S.~S., {Laughlin}, G., {et~al.} 2017, \aj, 153, 208

\bibitem[{{Connolly}(2009)}]{Connolly2009}
{Connolly}, J.~A.~D. 2009, Geochemistry, Geophysics, Geosystems, 10, Q10014

\bibitem[{{Crida} {et~al.}(2018){Crida}, {Ligi}, {Dorn}, \&
  {Lebreton}}]{Crida2018}
{Crida}, A., {Ligi}, R., {Dorn}, C., \& {Lebreton}, Y. 2018, ArXiv e-prints

\bibitem[{{Dawson} \& {Fabrycky}(2010)}]{Dawson2010}
{Dawson}, R.~I. \& {Fabrycky}, D.~C. 2010, \apj, 722, 937

\bibitem[{{de Mooij} {et~al.}(2014){de Mooij}, {L{\'o}pez-Morales},
  {Karjalainen}, {Hrudkova}, \& {Jayawardhana}}]{deMooij2014}
{de Mooij}, E.~J.~W., {L{\'o}pez-Morales}, M., {Karjalainen}, R., {Hrudkova},
  M., \& {Jayawardhana}, R. 2014, \apjl, 797, L21

\bibitem[{{Demory} {et~al.}(2015){Demory}, {Ehrenreich}, {Queloz}, {Seager},
  {Gilliland}, {Chaplin}, {Proffitt}, {Gillon}, {G{\"u}nther}, {Benneke},
  {Dumusque}, {Lovis}, {Pepe}, {S{\'e}gransan}, {Triaud}, \&
  {Udry}}]{Demory2015}
{Demory}, B.-O., {Ehrenreich}, D., {Queloz}, D., {et~al.} 2015, \mnras, 450,
  2043

\bibitem[{{Demory} {et~al.}(2016{\natexlab{a}}){Demory}, {Gillon}, {de Wit},
  {Madhusudhan}, {Bolmont}, {Heng}, {Kataria}, {Lewis}, {Hu}, {Krick},
  {Stamenkovi{\'c}}, {Benneke}, {Kane}, \& {Queloz}}]{Demory2016b}
{Demory}, B.-O., {Gillon}, M., {de Wit}, J., {et~al.} 2016{\natexlab{a}}, \nat,
  532, 207

\bibitem[{{Demory} {et~al.}(2011){Demory}, {Gillon}, {Deming}, {Valencia},
  {Seager}, {Benneke}, {Lovis}, {Cubillos}, {Harrington}, {Stevenson}, {Mayor},
  {Pepe}, {Queloz}, {S{\'e}gransan}, \& {Udry}}]{demory2011}
{Demory}, B.-O., {Gillon}, M., {Deming}, D., {et~al.} 2011, \aap, 533, A114

\bibitem[{{Demory} {et~al.}(2016{\natexlab{b}}){Demory}, {Gillon},
  {Madhusudhan}, \& {Queloz}}]{Demory2016a}
{Demory}, B.-O., {Gillon}, M., {Madhusudhan}, N., \& {Queloz}, D.
  2016{\natexlab{b}}, \mnras, 455, 2018

\bibitem[{{Demory} {et~al.}(2012){Demory}, {Gillon}, {Seager}, {Benneke},
  {Deming}, \& {Jackson}}]{Demory2012}
{Demory}, B.-O., {Gillon}, M., {Seager}, S., {et~al.} 2012, \apjl, 751, L28

\bibitem[{{Dorn} {et~al.}(2017{\natexlab{a}}){Dorn}, {Hinkel}, \&
  {Venturini}}]{Dorn2017}
{Dorn}, C., {Hinkel}, N.~R., \& {Venturini}, J. 2017{\natexlab{a}}, \aap, 597,
  A38

\bibitem[{{Dorn} {et~al.}(2017{\natexlab{b}}){Dorn}, {Venturini}, {Khan},
  {Heng}, {Alibert}, {Helled}, {Rivoldini}, \& {Benz}}]{Dorn2017a}
{Dorn}, C., {Venturini}, J., {Khan}, A., {et~al.} 2017{\natexlab{b}}, \aap,
  597, A37

\bibitem[{{Dragomir} {et~al.}(2013){Dragomir}, {Matthews}, {Eastman},
  {Cameron}, {Howard}, {Guenther}, {Kuschnig}, {Moffat}, {Rowe}, {Rucinski},
  {Sasselov}, \& {Weiss}}]{Dragomir2013_HD976}
{Dragomir}, D., {Matthews}, J.~M., {Eastman}, J.~D., {et~al.} 2013, \apjl, 772,
  L2

\bibitem[{{Dragomir} {et~al.}(2014){Dragomir}, {Matthews}, {Winn}, \&
  {Rowe}}]{Dragomir2013}
{Dragomir}, D., {Matthews}, J.~M., {Winn}, J.~N., \& {Rowe}, J.~F. 2014, in IAU
  Symposium, Vol. 293, IAU Symposium, ed. N.~{Haghighipour}, 52--57

\bibitem[{{Dumusque} {et~al.}(2011{\natexlab{a}}){Dumusque}, {Lovis},
  {S{\'e}gransan}, {Mayor}, {Udry}, {Benz}, {Bouchy}, {Lo Curto}, {Mordasini},
  {Pepe}, {Queloz}, {Santos}, \& {Naef}}]{Dumusque2011c}
{Dumusque}, X., {Lovis}, C., {S{\'e}gransan}, D., {et~al.} 2011{\natexlab{a}},
  \aap, 535, A55

\bibitem[{{Dumusque} {et~al.}(2011{\natexlab{b}}){Dumusque}, {Udry}, {Lovis},
  {Santos}, \& {Monteiro}}]{Dumusque2011a}
{Dumusque}, X., {Udry}, S., {Lovis}, C., {Santos}, N.~C., \& {Monteiro},
  M.~J.~P.~F.~G. 2011{\natexlab{b}}, \aap, 525, A140

\bibitem[{{Eastman} {et~al.}(2013){Eastman}, {Gaudi}, \& {Agol}}]{Eastman2013}
{Eastman}, J., {Gaudi}, B.~S., \& {Agol}, E. 2013, \pasp, 125, 83

\bibitem[{{Eastman} {et~al.}(2010){Eastman}, {Siverd}, \&
  {Gaudi}}]{Eastman2010}
{Eastman}, J., {Siverd}, R., \& {Gaudi}, B.~S. 2010, \pasp, 122, 935

\bibitem[{{Eaton} {et~al.}(2003){Eaton}, {Henry}, \& {Fekel}}]{Eaton2003}
{Eaton}, J.~A., {Henry}, G.~W., \& {Fekel}, F.~C. 2003, in Astrophysics and
  Space Science Library, Vol. 288, Astrophysics and Space Science Library, ed.
  T.~D. {Oswalt}, 189

\bibitem[{{Egeland} {et~al.}(2017){Egeland}, {Soon}, {Baliunas}, {Hall},
  {Pevtsov}, \& {Bertello}}]{Egeland2017}
{Egeland}, R., {Soon}, W., {Baliunas}, S., {et~al.} 2017, \apj, 835, 25

\bibitem[{{Ehrenreich} {et~al.}(2012){Ehrenreich}, {Bourrier}, {Bonfils},
  {Lecavelier des Etangs}, {H{\'e}brard}, {Sing}, {Wheatley}, {Vidal-Madjar},
  {Delfosse}, {Udry}, {Forveille}, \& {Moutou}}]{Ehrenreich2012}
{Ehrenreich}, D., {Bourrier}, V., {Bonfils}, X., {et~al.} 2012, \aap, 547, A18

\bibitem[{{Elkins-Tanton}(2012)}]{Elkins2012}
{Elkins-Tanton}, L.~T. 2012, Annual Review of Earth and Planetary Sciences, 40,
  113

\bibitem[{{Endl} {et~al.}(2012){Endl}, {Robertson}, {Cochran}, {MacQueen},
  {Brugamyer}, {Caldwell}, {Wittenmyer}, {Barnes}, \& {Gullikson}}]{Endl2012}
{Endl}, M., {Robertson}, P., {Cochran}, W.~D., {et~al.} 2012, \apj, 759, 19

\bibitem[{{Erkaev} {et~al.}(2007){Erkaev}, {Kulikov}, {Lammer}, {Selsis},
  {Langmayr}, {Jaritz}, \& {Biernat}}]{Erkaev2007}
{Erkaev}, N.~V., {Kulikov}, Y.~N., {Lammer}, H., {et~al.} 2007, \aap, 472, 329

\bibitem[{{Esteves} {et~al.}(2017){Esteves}, {de Mooij}, {Jayawardhana},
  {Watson}, \& {de Kok}}]{Esteves2017}
{Esteves}, L.~J., {de Mooij}, E.~J.~W., {Jayawardhana}, R., {Watson}, C., \&
  {de Kok}, R. 2017, \aj, 153, 268

\bibitem[{{Evans} {et~al.}(2013){Evans}, {Pont}, {Sing}, {Aigrain}, {Barstow},
  {D{\'e}sert}, {Gibson}, {Heng}, {Knutson}, \& {Lecavelier des
  Etangs}}]{Evans2013}
{Evans}, T.~M., {Pont}, F., {Sing}, D.~K., {et~al.} 2013, \apjl, 772, L16

\bibitem[{{Fischer}(2017)}]{Fischer2017}
{Fischer}, D.~A. 2017, 55 Cancri (Copernicus): A Multi-planet System with a Hot
  Super-Earth and a Jupiter Analogue, ed. B.~Deeg~H. (Springer)

\bibitem[{{Fischer} {et~al.}(2008){Fischer}, {Marcy}, {Butler}, {Vogt},
  {Laughlin}, {Henry}, {Abouav}, {Peek}, {Wright}, {Johnson}, {McCarthy}, \&
  {Isaacson}}]{Fischer2008}
{Fischer}, D.~A., {Marcy}, G.~W., {Butler}, R.~P., {et~al.} 2008, \apj, 675,
  790

\bibitem[{{Foreman-Mackey} {et~al.}(2013){Foreman-Mackey}, {Hogg}, {Lang}, \&
  {Goodman}}]{Foreman2013}
{Foreman-Mackey}, D., {Hogg}, D.~W., {Lang}, D., \& {Goodman}, J. 2013, \pasp,
  125, 306

\bibitem[{{Fulton} {et~al.}(2017){Fulton}, {Petigura}, {Howard}, {Isaacson},
  {Marcy}, {Cargile}, {Hebb}, {Weiss}, {Johnson}, {Morton}, {Sinukoff},
  {Crossfield}, \& {Hirsch}}]{Fulton2017}
{Fulton}, B.~J., {Petigura}, E.~A., {Howard}, A.~W., {et~al.} 2017, \aj, 154,
  109

\bibitem[{{Gillon} {et~al.}(2012){Gillon}, {Demory}, {Benneke}, {Valencia},
  {Deming}, {Seager}, {Lovis}, {Mayor}, {Pepe}, {Queloz}, {S{\'e}gransan}, \&
  {Udry}}]{Gillon2012}
{Gillon}, M., {Demory}, B.-O., {Benneke}, B., {et~al.} 2012, \aap, 539, A28

\bibitem[{{Ginzburg} {et~al.}(2016){Ginzburg}, {Schlichting}, \&
  {Sari}}]{Ginzburg2016}
{Ginzburg}, S., {Schlichting}, H.~E., \& {Sari}, R. 2016, \apj, 825, 29

\bibitem[{{Gomes da Silva} {et~al.}(2014){Gomes da Silva}, {Santos}, {Boisse},
  {Dumusque}, \& {Lovis}}]{GomesdaSilva2014}
{Gomes da Silva}, J., {Santos}, N.~C., {Boisse}, I., {Dumusque}, X., \&
  {Lovis}, C. 2014, \aap, 566, A66

\bibitem[{{Hammond} \& {Pierrehumbert}(2017)}]{Hammond2017}
{Hammond}, M. \& {Pierrehumbert}, R.~T. 2017, \apj, 849, 152

\bibitem[{{Henry}(1995{\natexlab{a}})}]{Henry_1995b}
{Henry}, G.~W. 1995{\natexlab{a}}, in Astronomical Society of the Pacific
  Conference Series, Vol.~79, Robotic Telescopes. Current Capabilities, Present
  Developments, and Future Prospects for Automated Astronomy, ed. G.~W. {Henry}
  \& J.~A. {Eaton}, 44

\bibitem[{{Henry}(1995{\natexlab{b}})}]{Henry_1995a}
{Henry}, G.~W. 1995{\natexlab{b}}, in Astronomical Society of the Pacific
  Conference Series, Vol.~79, Robotic Telescopes. Current Capabilities, Present
  Developments, and Future Prospects for Automated Astronomy, ed. G.~W. {Henry}
  \& J.~A. {Eaton}, 37

\bibitem[{{Henry}(1999)}]{Henry_1999}
{Henry}, G.~W. 1999, \pasp, 111, 845

\bibitem[{{Henry} {et~al.}(2000){Henry}, {Baliunas}, {Donahue}, {Fekel}, \&
  {Soon}}]{Henry2000}
{Henry}, G.~W., {Baliunas}, S.~L., {Donahue}, R.~A., {Fekel}, F.~C., \& {Soon},
  W. 2000, \apj, 531, 415

\bibitem[{{Holman} {et~al.}(2006){Holman}, {Winn}, {Latham}, {O'Donovan},
  {Charbonneau}, {Bakos}, {Esquerdo}, {Hergenrother}, {Everett}, \&
  {P{\'a}l}}]{Holman2006}
{Holman}, M.~J., {Winn}, J.~N., {Latham}, D.~W., {et~al.} 2006, \apj, 652, 1715

\bibitem[{{Huitson} {et~al.}(2012){Huitson}, {Sing}, {Vidal-Madjar},
  {Ballester}, {Lecavelier des Etangs}, {D{\'e}sert}, \& {Pont}}]{Huitson2012}
{Huitson}, C.~M., {Sing}, D.~K., {Vidal-Madjar}, A., {et~al.} 2012, \mnras,
  422, 2477

\bibitem[{Ito {et~al.}(2015)Ito, Ikoma, Kawahara, Nagahara, Kawashima, \&
  Nakamoto}]{Ito2015}
Ito, Y., Ikoma, M., Kawahara, H., {et~al.} 2015, The Astrophysical Journal,
  801, 144

\bibitem[{{Lecavelier des Etangs}(2007)}]{Lecav2007}
{Lecavelier des Etangs}, A. 2007, \aap, 461, 1185

\bibitem[{{Lopez}(2017)}]{Lopez2017}
{Lopez}, E.~D. 2017, \mnras, 472, 245

\bibitem[{{L{\'o}pez-Morales} {et~al.}(2014){L{\'o}pez-Morales}, {Triaud},
  {Rodler}, {Dumusque}, {Buchhave}, {Harutyunyan}, {Hoyer}, {Alonso}, {Gillon},
  {Kaib}, {Latham}, {Lovis}, {Pepe}, {Queloz}, {Raymond}, {S{\'e}gransan},
  {Waldmann}, \& {Udry}}]{lopez2014}
{L{\'o}pez-Morales}, M., {Triaud}, A.~H.~M.~J., {Rodler}, F., {et~al.} 2014,
  \apjl, 792, L31

\bibitem[{{Lovis} {et~al.}(2011){Lovis}, {Dumusque}, {Santos}, {Bouchy},
  {Mayor}, {Pepe}, {Queloz}, {S{\'e}gransan}, \& {Udry}}]{Lovis2011b}
{Lovis}, C., {Dumusque}, X., {Santos}, N.~C., {et~al.} 2011, ArXiv e-prints

\bibitem[{{Madhusudhan} {et~al.}(2012){Madhusudhan}, {Lee}, \&
  {Mousis}}]{Madhusudhan2012}
{Madhusudhan}, N., {Lee}, K.~K.~M., \& {Mousis}, O. 2012, \apjl, 759, L40

\bibitem[{{Mamajek} \& {Hillenbrand}(2008)}]{Mamajek2008}
{Mamajek}, E.~E. \& {Hillenbrand}, L.~A. 2008, \apj, 687, 1264

\bibitem[{{Mandel} \& {Agol}(2002)}]{Mandel2002}
{Mandel}, K. \& {Agol}, E. 2002, \apjl, 580, L171

\bibitem[{{Marcy} {et~al.}(2002){Marcy}, {Butler}, {Fischer}, {Laughlin},
  {Vogt}, {Henry}, \& {Pourbaix}}]{Marcy2002}
{Marcy}, G.~W., {Butler}, R.~P., {Fischer}, D.~A., {et~al.} 2002, \apj, 581,
  1375

\bibitem[{{McArthur} {et~al.}(2004){McArthur}, {Endl}, {Cochran}, {Benedict},
  {Fischer}, {Marcy}, {Butler}, {Naef}, {Mayor}, {Queloz}, {Udry}, \&
  {Harrison}}]{McArthur2004}
{McArthur}, B.~E., {Endl}, M., {Cochran}, W.~D., {et~al.} 2004, \apjl, 614, L81

\bibitem[{{Meunier} {et~al.}(2010{\natexlab{a}}){Meunier}, {Desort}, \&
  {Lagrange}}]{Meunier2010a}
{Meunier}, N., {Desort}, M., \& {Lagrange}, A.-M. 2010{\natexlab{a}}, \aap,
  512, A39

\bibitem[{{Meunier} {et~al.}(2010{\natexlab{b}}){Meunier}, {Lagrange}, \&
  {Desort}}]{Meunier2010b}
{Meunier}, N., {Lagrange}, A.-M., \& {Desort}, M. 2010{\natexlab{b}}, \aap,
  519, A66

\bibitem[{{Motalebi} {et~al.}(2015){Motalebi}, {Udry}, {Gillon}, {Lovis},
  {S{\'e}gransan}, {Buchhave}, {Demory}, {Malavolta}, {Dressing}, {Sasselov},
  {Rice}, {Charbonneau}, {Collier Cameron}, {Latham}, {Molinari}, {Pepe},
  {Affer}, {Bonomo}, {Cosentino}, {Dumusque}, {Figueira}, {Fiorenzano},
  {Gettel}, {Harutyunyan}, {Haywood}, {Johnson}, {Lopez}, {Lopez-Morales},
  {Mayor}, {Micela}, {Mortier}, {Nascimbeni}, {Philips}, {Piotto}, {Pollacco},
  {Queloz}, {Sozzetti}, {Vanderburg}, \& {Watson}}]{Motalebi2015}
{Motalebi}, F., {Udry}, S., {Gillon}, M., {et~al.} 2015, \aap, 584, A72

\bibitem[{{Mugrauer} {et~al.}(2006){Mugrauer}, {Neuh{\"a}user}, {Mazeh},
  {Guenther}, {Fern{\'a}ndez}, \& {Broeg}}]{Mugrauer2006}
{Mugrauer}, M., {Neuh{\"a}user}, R., {Mazeh}, T., {et~al.} 2006, Astronomische
  Nachrichten, 327, 321

\bibitem[{{Nelson} {et~al.}(2014){Nelson}, {Ford}, {Wright}, {Fischer}, {von
  Braun}, {Howard}, {Payne}, \& {Dindar}}]{Nelson2014}
{Nelson}, B.~E., {Ford}, E.~B., {Wright}, J.~T., {et~al.} 2014, \mnras

\bibitem[{{Noyes} {et~al.}(1984){Noyes}, {Hartmann}, {Baliunas}, {Duncan}, \&
  {Vaughan}}]{Noyes1984}
{Noyes}, R.~W., {Hartmann}, L.~W., {Baliunas}, S.~L., {Duncan}, D.~K., \&
  {Vaughan}, A.~H. 1984, \apj, 279, 763

\bibitem[{{Pont} {et~al.}(2006){Pont}, {Zucker}, \& {Queloz}}]{pont2006}
{Pont}, F., {Zucker}, S., \& {Queloz}, D. 2006, \mnras, 373, 231

\bibitem[{{Ridden-Harper} {et~al.}(2016){Ridden-Harper}, {Snellen}, {Keller},
  {de Kok}, {Di Gloria}, {Hoeijmakers}, {Brogi}, {Fridlund}, {Vermeersen}, \&
  {van Westrenen}}]{RiddenHarper2016}
{Ridden-Harper}, A.~R., {Snellen}, I.~A.~G., {Keller}, C.~U., {et~al.} 2016,
  \aap, 593, A129

\bibitem[{{Salz} {et~al.}(2016){Salz}, {Schneider}, {Czesla}, \&
  {Schmitt}}]{Salz2016a}
{Salz}, M., {Schneider}, P.~C., {Czesla}, S., \& {Schmitt}, J.~H.~M.~M. 2016,
  \aap, 585, L2

\bibitem[{{Seager} {et~al.}(2007){Seager}, {Kuchner}, {Hier-Majumder}, \&
  {Militzer}}]{Seager2007}
{Seager}, S., {Kuchner}, M., {Hier-Majumder}, C.~A., \& {Militzer}, B. 2007,
  \apj, 669, 1279

\bibitem[{{Sing} {et~al.}(2013){Sing}, {Lecavelier des Etangs}, {Fortney},
  {Burrows}, {Pont}, {Wakeford}, {Ballester}, {Nikolov}, {Henry}, {Aigrain},
  {Deming}, {Evans}, {Gibson}, {Huitson}, {Knutson}, {Showman}, {Vidal-Madjar},
  {Wilson}, {Williamson}, \& {Zahnle}}]{Sing2013}
{Sing}, D.~K., {Lecavelier des Etangs}, A., {Fortney}, J.~J., {et~al.} 2013,
  \mnras, 436, 2956

\bibitem[{{Sing} {et~al.}(2008){Sing}, {Vidal-Madjar}, {D{\'e}sert},
  {Lecavelier des Etangs}, \& {Ballester}}]{Sing2008a}
{Sing}, D.~K., {Vidal-Madjar}, A., {D{\'e}sert}, J.-M., {Lecavelier des
  Etangs}, A., \& {Ballester}, G. 2008, \apj, 686, 658

\bibitem[{{Tsiaras} {et~al.}(2016){Tsiaras}, {Rocchetto}, {Waldmann}, {Venot},
  {Varley}, {Morello}, {Damiano}, {Tinetti}, {Barton}, {Yurchenko}, \&
  {Tennyson}}]{Tsiaras2016}
{Tsiaras}, A., {Rocchetto}, M., {Waldmann}, I.~P., {et~al.} 2016, \apj, 820, 99

\bibitem[{{Valenti} \& {Fischer}(2005)}]{Valenti2005}
{Valenti}, J.~A. \& {Fischer}, D.~A. 2005, \apjs, 159, 141

\bibitem[{{Vazan} {et~al.}(2013){Vazan}, {Kovetz}, {Podolak}, \&
  {Helled}}]{Vazan2013}
{Vazan}, A., {Kovetz}, A., {Podolak}, M., \& {Helled}, R. 2013, \mnras, 434,
  3283

\bibitem[{{von Braun} {et~al.}(2011){von Braun}, {Boyajian}, {ten Brummelaar},
  {Kane}, {van Belle}, {Ciardi}, {Raymond}, {L{\'o}pez-Morales}, {McAlister},
  {Schaefer}, {Ridgway}, {Sturmann}, {Sturmann}, {White}, {Turner},
  {Farrington}, \& {Goldfinger}}]{vonbraun2011}
{von Braun}, K., {Boyajian}, T.~S., {ten Brummelaar}, T.~A., {et~al.} 2011,
  \apj, 740, 49

\bibitem[{{Wilson} {et~al.}(2015){Wilson}, {Sing}, {Nikolov}, {Lecavelier des
  Etangs}, {Pont}, {Fortney}, {Ballester}, {L{\'o}pez-Morales}, {D{\'e}sert},
  \& {Vidal-Madjar}}]{Wilson2015}
{Wilson}, P.~A., {Sing}, D.~K., {Nikolov}, N., {et~al.} 2015, \mnras, 450, 192

\bibitem[{{Winn} {et~al.}(2009){Winn}, {Holman}, {Henry}, {Torres}, {Fischer},
  {Johnson}, {Marcy}, {Shporer}, \& {Mazeh}}]{Winn2009d}
{Winn}, J.~N., {Holman}, M.~J., {Henry}, G.~W., {et~al.} 2009, \apj, 693, 794

\bibitem[{{Winn} {et~al.}(2011){Winn}, {Matthews}, {Dawson}, {Fabrycky},
  {Holman}, {Kallinger}, {Kuschnig}, {Sasselov}, {Dragomir}, {Guenther},
  {Moffat}, {Rowe}, {Rucinski}, \& {Weiss}}]{Winn2011}
{Winn}, J.~N., {Matthews}, J.~M., {Dawson}, R.~I., {et~al.} 2011, \apjl, 737,
  L18

\bibitem[{{Yee} {et~al.}(2017){Yee}, {Petigura}, \& {von Braun}}]{Yee2017}
{Yee}, S.~W., {Petigura}, E.~A., \& {von Braun}, K. 2017, \apj, 836, 77

\bibitem[{{Zechmeister} \& {K{\"u}rster}(2009)}]{Zechmeister2009}
{Zechmeister}, M. \& {K{\"u}rster}, M. 2009, \aap, 496, 577

\bibitem[{{Zhang} \& {Showman}(2017)}]{Zhang2016}
{Zhang}, X. \& {Showman}, A.~P. 2017, \apj, 836, 73

\end{thebibliography}

\begin{appendix}
\label{apn:appendix}

\begin{deluxetable}{cccccccc}
\tabletypesize{\small}
\rotate
\tablewidth{19cm}
\tablecaption{SUMMARY OF T8 0.80-m APT OBSERVATIONS OF 55 CANCRI}
\label{tab:photom}
\tablehead{

\colhead{Observing} & \colhead{Mean Julian date} & \colhead{} & \colhead{Seasonal mean} & \colhead{Seasonal mean} & \colhead{Seasonal mean} & \colhead{Period} & \colhead{Amplitude} \\

\colhead{season} & \colhead{(HJD$-$2,400,000)} & \colhead{$N_{obs}$} & \colhead{($P-C2_{by}$)} & \colhead{($P-C3_{by}$)} & \colhead{($C3-C2_{by}$)} & \colhead{(days)} & \colhead{(mag)}  \\

\colhead{(1)} & \colhead{(2)} & \colhead{(3)} & \colhead{(4)} & \colhead{(5)} & \colhead{(6)} & \colhead{(7)} & \colhead{(8)}

}
\startdata
2000-2001 & 2451972.7 &  81 & $0.07336\pm0.00012$ & $-0.69247\pm0.00011$ & $0.76585\pm0.00011$ &     \nodata    &       \nodata       \\
2001-2002 & 2452302.2 & 116 & $0.07306\pm0.00010$ & $-0.69297\pm0.00010$ & $0.76604\pm0.00008$ &     \nodata    &       \nodata       \\
2002-2003 & 2452671.5 & 135 & $0.07315\pm0.00011$ & $-0.69314\pm0.00012$ & $0.76629\pm0.00010$ &     \nodata    &       \nodata       \\
2003-2004 & 2453025.7 & 150 & $0.07232\pm0.00012$ & $-0.69399\pm0.00011$ & $0.76631\pm0.00010$ &     \nodata    &       \nodata       \\
2004-2005 & 2453388.4 & 157 & $0.07284\pm0.00018$ & $-0.69326\pm0.00017$ & $0.76609\pm0.00008$ & $39.34\pm0.85$ & $0.00072\pm0.00016$ \\
2005-2006 & 2453742.2 & 145 & $0.07197\pm0.00014$ & $-0.69425\pm0.00016$ & $0.76623\pm0.00009$ & $46.23\pm1.38$ & $0.00065\pm0.00017$ \\
2006-2007 & 2454118.8 & 132 & $0.07305\pm0.00020$ & $-0.69337\pm0.00019$ & $0.76643\pm0.00007$ & $41.67\pm0.84$ & $0.00085\pm0.00017$ \\
2007-2008 & 2454512.8 & 636 & $0.07250\pm0.00006$ & $-0.69399\pm0.00007$ & $0.76649\pm0.00004$ &     \nodata    &       \nodata       \\
2008-2009 & 2454849.2 &  61 & $0.07381\pm0.00020$ & $-0.69248\pm0.00021$ & $0.76628\pm0.00012$ & $40.44\pm0.75$ & $0.00140\pm0.00025$ \\
2009-2010 & 2455211.3 &  89 & $0.07343\pm0.00013$ & $-0.69301\pm0.00013$ & $0.76644\pm0.00010$ & $34.94\pm1.07$ & $0.00065\pm0.00022$ \\
2010-2011 & 2455581.5 & 112 & $0.07440\pm0.00009$ & $-0.69226\pm0.00011$ & $0.76665\pm0.00008$ &     \nodata    &       \nodata       \\
2011-2012 & 2455959.9 &  99 & $0.07436\pm0.00010$ & $-0.69200\pm0.00011$ & $0.76637\pm0.00008$ &     \nodata    &       \nodata       \\
2012-2013 & 2456318.9 &  74 & $0.07436\pm0.00009$ & $-0.69244\pm0.00011$ & $0.76681\pm0.00010$ &     \nodata    &       \nodata       \\
2013-2014 & 2456680.9 &  85 & $0.07405\pm0.00010$ & $-0.69227\pm0.00010$ & $0.76631\pm0.00006$ &     \nodata    &       \nodata       \\
2014-2015 & 2457035.9 &  53 & $0.07431\pm0.00025$ & $-0.69198\pm0.00031$ & $0.76629\pm0.00015$ &     \nodata    &       \nodata       \\
2015-2016 & 2457413.4 &  70 & $0.07380\pm0.00034$ & $-0.69267\pm0.00030$ & $0.76646\pm0.00016$ &     \nodata    &       \nodata       \\
2016-2017 & 2457755.3 &  48 & $0.07385\pm0.00032$ & $-0.69299\pm0.00029$ & $0.76681\pm0.00018$ & $38.45\pm1.36$ & $0.00132\pm0.00030$ \\
\enddata
\end{deluxetable}

\begin{table*}
\scriptsize
\caption{Parameters probed by the MCMC used to fit the RV measurements of 55\,Cnc. The maximum likelihood solution (Max(Like)), the median (Med), mode (Mod) and standard deviation (Std) of the posterior distribution for each parameter is shown, as well as the 68.3\% (CI(15.85),CI(84.15)) and 95.45\% (CI(2.275),CI(97.725)) confidence intervals. The prior for each parameter can be of type: $\mathcal{U}$: uniform, $\mathcal{N}$: normal, $\mathcal{SN}$:split normal, or $\mathcal{TN}$:truncated normal}.  
\label{tab:55CANCRI_tab-mcmc-Probed_params}
\def\arraystretch{1.5}
\begin{center}
\begin{tabular}{lcclcccccccc}
\hline
\hline
Param. & Units & Max(Like) & Med & Mod &Std & CI(15.85) & CI(84.15) &CI(2.275) & CI(97.725) & Prior\\
\hline
\multicolumn{11}{c}{ \bf Likelihood}\\
\hline
$\log{(\rm Post})$&               & -2273.450488& -2285.842126& -2284.433522&     4.255952& -2290.966384& -2281.311469& -2297.110579& -2277.307845&               \\ 
$\log{(\rm Like)}$&               & -2291.046969& -2302.472513& -2302.414228&     4.194558& -2307.394292& -2298.010475& -2313.297187& -2293.907193&               \\ 
$\log{(\rm Prior)}$&               &    17.596481&    16.873550&    17.204741&     0.721226&    15.862038&    17.365749&    14.233444&    17.547586&               \\ 
\hline 
M$_{\star}$    &[M$_{\odot}$]  &     0.892695&     0.904364&     0.906670&     0.013292&     0.889584&     0.919407&     0.874257&     0.935560&$\mathcal{N}(0.905,0.015)$\\ 
\hline 
$\sigma_{HARPN}$&[m\,s$^{-1}$]  &     1.06&     1.34&     1.18&     0.26&     1.09&     1.66&     0.89&     2.10&$\mathcal{U}$  \\ 
$\sigma_{HARPS}$&[m\,s$^{-1}$]  &     0.42&     0.67&     0.46&     0.39&     0.26&     1.13&     0.03&     1.80&$\mathcal{U}$  \\ 
$\sigma_{HRS}$ &[m\,s$^{-1}$]  &     2.28&     3.52&     3.62&     0.68&     2.72&     4.21&     1.52&     4.87&$\mathcal{U}$  \\ 
$\sigma_{KECK}$&[m\,s$^{-1}$]  &     2.91&     3.35&     3.30&     0.21&     3.12&     3.59&     2.86&     3.85&$\mathcal{U}$  \\ 
$\sigma_{LICK}$&[m\,s$^{-1}$]  &     5.44&     5.81&     5.76&     0.37&     5.36&     6.22&     4.93&     6.58&$\mathcal{U}$  \\ 
$\sigma_{SOPHIE}$&[m\,s$^{-1}$]  &     1.69&     1.95&     1.93&     0.28&     1.66&     2.29&     1.39&     2.69&$\mathcal{U}$  \\ 
$\sigma_{TULL}$&[m\,s$^{-1}$]  &     3.59&     3.89&     3.82&     0.35&     3.53&     4.31&     3.17&     4.76&$\mathcal{U}$  \\ 
$\sigma_{JIT}$ &[m\,s$^{-1}$]  &     3.81&     2.81&     2.76&     0.90&     1.66&     3.71&     0.39&     4.40&$\mathcal{U}$  \\ 
\hline 
$\gamma_{HARPN}$&[m\,s$^{-1}$]  & 27451.45& 27451.91& 27451.85&     0.89& 27450.84& 27452.89& 27449.86& 27453.93&$\mathcal{U}$  \\ 
$\gamma_{HARPS}$&[m\,s$^{-1}$]  & 27468.87& 27469.02& 27468.75&     0.82& 27468.13& 27469.97& 27467.21& 27470.93&$\mathcal{U}$  \\ 
$\gamma_{HRS}$ &[m\,s$^{-1}$]  & 28396.92& 28397.31& 28397.27&     0.79& 28396.39& 28398.21& 28395.53& 28399.10&$\mathcal{U}$  \\ 
$\gamma_{KECK}$&[m\,s$^{-1}$]  &   -40.91&   -40.89&   -40.95&     0.40&   -41.34&   -40.42&   -41.75&   -39.92&$\mathcal{U}$  \\ 
$\gamma_{LICK}$&[m\,s$^{-1}$]  &     3.43&     3.74&     3.60&     0.41&     3.28&     4.22&     2.83&     4.72&$\mathcal{U}$  \\ 
$\gamma_{SOPHIE}$&[m\,s$^{-1}$]  & 27437.78& 27438.00& 27437.85&     0.81& 27437.10& 27438.96& 27436.19& 27439.84&$\mathcal{U}$  \\ 
$\gamma_{TULL}$&[m\,s$^{-1}$]  &-22571.04&-22571.01&-22571.27&     0.57&-22571.66&-22570.35&-22572.30&-22569.68&$\mathcal{U}$  \\ 
\hline 
\end{tabular}
\end{center}
\end{table*}

\begin{table*}
\scriptsize
\caption{Continuation of Table \ref{tab:55CANCRI_tab-mcmc-Probed_params}}.  
\label{tab:55CANCRI_tab-mcmc-Probed_params2}
\def\arraystretch{1.5}
\begin{center}
\begin{tabular}{lcclcccccccc}
\hline
\hline
Param. & Units & Max(Like) & Med & Mod &Std & CI(15.85) & CI(84.15) &CI(2.275) & CI(97.725) & Prior\\
\hline
\multicolumn{11}{c}{ \bf Likelihood}\\
\hline 
$\log{(P)}$    &[d]            &     1.165883&     1.165885&     1.165885&     0.000002&     1.165883&     1.165887&     1.165881&     1.165889&$\mathcal{U}$  \\ 
$\log{(K)}$    &[m\,s$^{-1}$]  &     1.85&     1.85&     1.85&     0.00&     1.85&     1.85&     1.85&     1.86&$\mathcal{U}$  \\ 
$\sqrt{e}.\cos{\omega}$&               &     0.022977&     0.031298&     0.044307&     0.031656&    -0.009855&     0.064623&    -0.043345&     0.086918&$\mathcal{U}$  \\ 
$\sqrt{e}.\sin{\omega}$&               &    -0.076941&    -0.006118&    -0.005375&     0.030482&    -0.041314&     0.030238&    -0.066334&     0.058214&$\mathcal{U}$  \\ 
$\lambda_{0}$  &[deg]          &   198.337280&   198.236638&   198.254369&     0.170587&   198.040883&   198.429919&   197.842096&   198.617500&$\mathcal{U}$  \\ 
\hline 
$\log{(P)}$    &[d]            &     1.647371&     1.647372&     1.647365&     0.000036&     1.647330&     1.647413&     1.647286&     1.647457&$\mathcal{U}$  \\ 
$\log{(K)}$    &[m\,s$^{-1}$]  &     0.99&     1.00&     0.99&     0.01&     0.99&     1.00&     0.98&     1.01&$\mathcal{U}$  \\ 
$\sqrt{e}.\cos{\omega}$&               &     0.145067&     0.139330&     0.154716&     0.082223&     0.025951&     0.214780&    -0.079089&     0.266234&$\mathcal{U}$  \\ 
$\sqrt{e}.\sin{\omega}$&               &    -0.021607&     0.007209&     0.019499&     0.083988&    -0.092205&     0.104252&    -0.174251&     0.175082&$\mathcal{U}$  \\ 
$\lambda_{0}$  &[deg]          &   152.107645&   152.036764&   151.667729&     1.009441&   150.913959&   153.195011&   149.746100&   154.276811&$\mathcal{U}$  \\ 
\hline 
$\log{(P)}$    &[d]            &     2.414600&     2.414770&     2.414529&     0.000433&     2.414285&     2.415259&     2.413794&     2.415791&$\mathcal{U}$  \\ 
$\log{(K)}$    &[m\,s$^{-1}$]  &     0.73&     0.71&     0.71&     0.02&     0.69&     0.73&     0.66&     0.75&$\mathcal{U}$  \\ 
$\sqrt{e}.\cos{\omega}$&               &     0.028362&    -0.029096&    -0.132560&     0.133065&    -0.187318&     0.130365&    -0.294743&     0.249554&$\mathcal{U}$  \\ 
$\sqrt{e}.\sin{\omega}$&               &    -0.270494&    -0.225158&    -0.278934&     0.113215&    -0.322547&    -0.067598&    -0.392342&     0.091226&$\mathcal{U}$  \\ 
$\lambda_{0}$  &[deg]          &   106.069275&   102.615196&   103.432101&     2.715291&    99.345207&   105.648564&    96.345607&   108.304627&$\mathcal{U}$  \\ 
\hline 
$\log{(P)}$    &[d]            &     3.580323&     3.582340&     3.581415&     0.007485&     3.573450&     3.590936&     3.565279&     3.598119&$\mathcal{U}$  \\ 
$\log{(K)}$    &[m\,s$^{-1}$]  &     1.18&     1.18&     1.19&     0.05&     1.13&     1.22&     1.06&     1.27&$\mathcal{U}$  \\ 
$\sqrt{e}.\cos{\omega}$&               &    -0.391429&    -0.400855&    -0.384878&     0.043763&    -0.455229&    -0.354370&    -0.499443&    -0.309570&$\mathcal{U}$  \\ 
$\sqrt{e}.\sin{\omega}$&               &     0.049300&     0.038499&     0.065933&     0.094395&    -0.079834&     0.137095&    -0.186755&     0.227198&$\mathcal{U}$  \\ 
$\lambda_{0}$  &[deg]          &   122.695667&   124.709302&   126.275021&     5.071957&   119.011467&   130.603288&   114.471907&   137.366403&$\mathcal{U}$  \\ 
\hline 
$\log{(P)}$    &[d]            &     3.749135&     3.746185&     3.745006&     0.006227&     3.739227&     3.753435&     3.733291&     3.760178&$\mathcal{U}$  \\ 
$\log{(K)}$    &[m\,s$^{-1}$]  &     1.58&     1.59&     1.59&     0.01&     1.57&     1.60&     1.56&     1.61&$\mathcal{U}$  \\ 
$\sqrt{e}.\cos{\omega}$&               &     0.134776&     0.128560&     0.127138&     0.043737&     0.081188&     0.180318&     0.025802&     0.223687&$\mathcal{U}$  \\ 
$\sqrt{e}.\sin{\omega}$&               &    -0.328459&    -0.334039&    -0.346717&     0.031095&    -0.368411&    -0.298759&    -0.403605&    -0.261516&$\mathcal{U}$  \\ 
$\lambda_{0}$  &[deg]          &     7.523549&     8.373486&     7.475054&     3.442002&     4.614703&    12.050153&     0.232590&    15.802723&$\mathcal{U}$  \\ 
\hline 
$P$            &[d]            &     0.736546&     0.736547&     0.736547&     0.000001&     0.736546&     0.736549&     0.736545&     0.736550&$\mathcal{N}(0.7365462780,1.8477e-06)$\\ 
$\log{(K)}$    &[m\,s$^{-1}$]  &     0.77&     0.78&     0.78&     0.01&     0.76&     0.80&     0.74&     0.81&$\mathcal{U}$  \\ 
$\sqrt{e}.\cos{\omega}$&               &     0.065426&     0.012710&    -0.019946&     0.079210&    -0.083201&     0.105674&    -0.160391&     0.174811&$\mathcal{U}$  \\ 
$\sqrt{e}.\sin{\omega}$&               &     0.234745&     0.192120&     0.220349&     0.093920&     0.063752&     0.268215&    -0.076741&     0.331372&$\mathcal{U}$  \\ 
$T_C$          &[d]            & 55733.005597& 55733.005980& 55733.006299&     0.001265& 55733.004556& 55733.007423& 55733.003120& 55733.008795&$\mathcal{N}(55733.0058594,0.0014648)$\\ 
\hline 
\end{tabular}
\end{center}
\end{table*}

\end{appendix}

\end{document}